\documentstyle[twocolumn,psfig,rotate]{mn}
\begin {document}

\title{A radio and optical study of Molonglo radio sources}

\author[C.H. Ishwara-Chandra et al.] 
{C.H.  Ishwara-Chandra$^{1,2}$\thanks{E-mail: ishwar@rri.res.in}, 
D.J. Saikia$^{1}$\thanks{E-mail: djs@ncra.tifr.res.in}, 
P.J. McCarthy$^{3}$\thanks{E-mail: pmc2@ociw.edu} and 
W.J.M. van Breugel$^{4}$\thanks{E-mail:wjm@iigp.llnl.gov.edu} \\ 
$^{1}$ National Centre for Radio Astrophysics, TIFR, Post Bag 3, 
Ganeshkhind, Pune 411 007, India \\ 
$^{2}$ Technical Physics Division, ISRO Satellite Center, Airport Road, 
Bangalore 560 017, India \\ 
$^{3}$ The Observatories of the Carnegie Institution of
Washington, 813 Santa Barbara St, Pasadena, CA 91101, USA \\ 
$^{4}$ Institute of Geophysics and Planetary Physics, 
Lawrence Livermore National Laboratory, L-413, Livermore, CA 94550, USA} 

\date{Received: }

\maketitle

\begin{abstract} 

We present multi-wavelength radio observations with the
Very Large Array and narrow- and broad-band optical observations with the
2.5m telescope at the Las Campanas Observatory, of a well-defined sample
of high-luminosity Fanaroff-Riley class II radio galaxies and quasars, 
selected from the Molonglo Reference Catalogue 1-Jy sample.
These observations were carried out as part of a programme to investigate
the effects of orientation and environment on some of the observed properties
of these sources. We examine
the dependence of the Liu-Pooley relationship, which shows that radio
lobes with flatter radio spectra are less depolarized, on size,
identification and redshift, and show that it is significantly stronger
for smaller sources, with the strength of the relationship being similar
for both radio galaxies and quasars. In addition to Doppler effects, there
appears to be intrinsic differences between the lobes on opposite sides.  
We discuss the asymmetry in brightness and location of the hotspots, and
present estimates of the ages and velocities from matched resolution
observations at L- and C-bands. The narrow- and broad-band optical 
images of some of these sources were made to study their environments 
and correlate with the symmetry parameters. 
Extended emission-line region is seen in a quasar and in four of the objects,
possible companion galaxies are seen close to the radio axis.

\end{abstract}


\section{Introduction} 

There has been a reasonable degree of evidence in support of the unified
scheme for powerful radio galaxies and quasars, according to which these
sources are intrinsically similar, but appear to be different because they
are oriented at different angles to the line of sight (Scheuer 1987;
Barthel 1989; Antonucci 1993; Urry \& Padovani 1995).  When the object is
viewed within about 45$^\circ$ to the line of sight, it appears as a
quasar, while it appears as a radio galaxy when it is oriented at larger
angles. The quasars tend to have more prominent radio cores and and
one-sided jets compared to radio galaxies, consistent with the ideas of
the unified scheme.  The prominence and apparent one-sidedness of jets is
attributed to relativistic beaming of the radio emission in sources
inclined at small angles to the line of sight.  One of the strongest
pieces of evidence in favour of the relativistic beaming hypothesis for
the apparent asymmetry of the jets comes from the the discovery by Laing
(1988) and Garrington et al. (1988)  that double radio sources depolarize
less rapidly on the side with the radio jet than on the opposite side. The
approaching side is seen through less of the depolarizing medium, and the
Laing-Garrington effect can be understood as an orientation effect (cf.  
Garrington, Conway \& Leahy 1991; Garrington \& Conway 1991). In addition
to the effects of orientation, the depolarization of the lobes could also
be affected by any asymmetry in the distribution of gas in the vicinity of
the radio source.  The possibility of an intrinsic asymmetry in the
distribution of gas was suspected from the fact that the lobe on the jet
side, which is approaching us, is often closer to the nucleus (Saikia
1981). An intrinsic asymmetry was demonstrated clearly by McCarthy, van
Breugel \& Kapahi (1991) who showed that there was invariably more
emission-line gas on the side of the source which is closer to the
nucleus. For a sample of 12 radio galaxies observed with the Very Large
Array (VLA), Pedelty et al. (1989) found the arm-length ratios to be
correlated with the amount of depolarization and emission line gas.

We have been making a systematic radio and optical study of a well-defined
sample of Fanaroff-Riley class II radio sources selected from the 408 MHz
Molonglo Reference Catalogue (McCarthy et al. 1990, 1996; Kapahi et al.
1998a,b; Baker et al. 1999), to investigate the effects of orientation and
environment on the observed symmetry parameters and depolarization
characteristics of the lobes of radio emission. The complete sample
consists of about 560 sources with S$_{408}\geq$0.95 Jy in the declination
range $-30^\circ \leq \delta -20^\circ$ and above a galactic latitude of
20$^\circ$. From this complete sample we compiled a list of 15 quasars and
a representative sample of 27 radio galaxies, which have similar size,
redshift and luminosity distributions, to study the effects of orientation
and environment in this class of objects. Sources were restricted to
those with a somewhat large angular size ($\geq 50 ''$), so that it would
also be possible to make further measurements at lower frequencies (610
and 327 MHz) with the Giant Metrewave Radio Telescope, with a reasonable
number of beamwidths along the source axes. Some of the angular sizes listed
in Ishwara-Chandra et al. (1998, hereinafter referred to as IC98), are incorrect, although
this has not affected the results presented in the paper. The sample of 42
sources along with the corrected sizes is listed in Appendix 1. 

\begin{table}
\centering
\caption{Observing log : Radio}
\begin{tabular}{l l l l l }
       &       &         &        &         \\
Array  & Obs.  &Obs.     & Band-  & Date of obs.  \\
Conf.  & band  &Freq.    & width  &       \\
       &       & MHz     & MHz    &        \\
       &       &         &        &         \\
BnA    & L     &1365   &  50 &  1995 Sep 20  \\
       & L     &1665   &  25 &                \\
CnB    & C     &4635   &  50 &  1996 Jan 20,31 \\
       & C     &4935   &  50 &                  \\        
DnC    & U     &14965  &  50 &  1997 Sep 15, 16  \\
DnC    & U     &14965  &  50 &  1997 Oct 3, 4, 12 \\
BnA    & X     &8447   &  25 &  1997 Feb 3  \\
\end{tabular}
\end{table}

Some of the results from our study have been presented earlier. The
polarization observations of the lobes at L- and C-bands and the effects
of the environment and orientation on the observed depolarization
characteristics of the lobes have been presented in 
IC98. These observations show that the
nearer lobe tends to be more depolarized and is also brighter, suggesting
that these objects are evolving in an asymmetric environment. In addition,
the effects of orientation are also marginally seen with the quasars
exhibiting a higher degree of depolarization asymmetry compared to the
radio galaxies. Radio observations of two new giant quasars found in our
sample, along with a study of the evolution of giant radio sources have
been presented by Ishwara-Chandra \& Saikia (1999). We have also
determined the spectra of the cores and hotspots and demonstrated that
they show the effects of relativistic motion, consistent with the unified
scheme.  We have used the measurements to estimate the velocity of advance
of the hotspots (Ishwara-Chandra \& Saikia 2000). In this paper, we 
describe the results of the radio and optical observations (Sections 2 and
3), and present a few typical images of the radio galaxies and quasars at the X- and
U-bands. We examine the Liu-Pooley effect (Liu \& Pooley 1991) and discuss
the brightness asymmetries of the hotspots (Section 3). We present the
estimates of ages and velocities in our sample using the matched
resolution observations at the L- and C-bands (Section 4). In Section 5,
the optical narrow- and broad-band observations and their results are
presented.  The conclusions are summarized in Section 6. We have defined
the spectral index $\alpha$ as $S_\nu \propto \nu^{-\alpha}$.

\section{Observations and analyses}

\subsection{Radio observations}

The observations were made with scaled arrays of the VLA at 1.4 and 1.7
GHz (L-band), 5 GHz (C-band) and 15 GHz (U-band)  with resolutions of
$\sim$5$^{\prime\prime}$, while observations at 8 GHz (X-band) have a
resolution of $\sim$1$^{\prime\prime}$. At 15 GHz, 14 quasars and 10 radio
galaxies were observed, while at 8 GHz only those sources in the RA range
03h to 13h were observed due to scheduling constraints. The observing log
is summarized in Table 1. All the data were calibrated and analyzed using
the National Radio Astronomy Observatory (NRAO) {\tt AIPS} package. 
The images at X band and U band were
corrected for primary beam attenuation. The U-band images have been
restored with the same resolutions as the L- and C-band images, which are
listed in Table 2.3 of IC98. However, no polarization information is 
available for sources observed in the X- and U-band.

\begin{figure*}
\hbox{
\psfig{figure=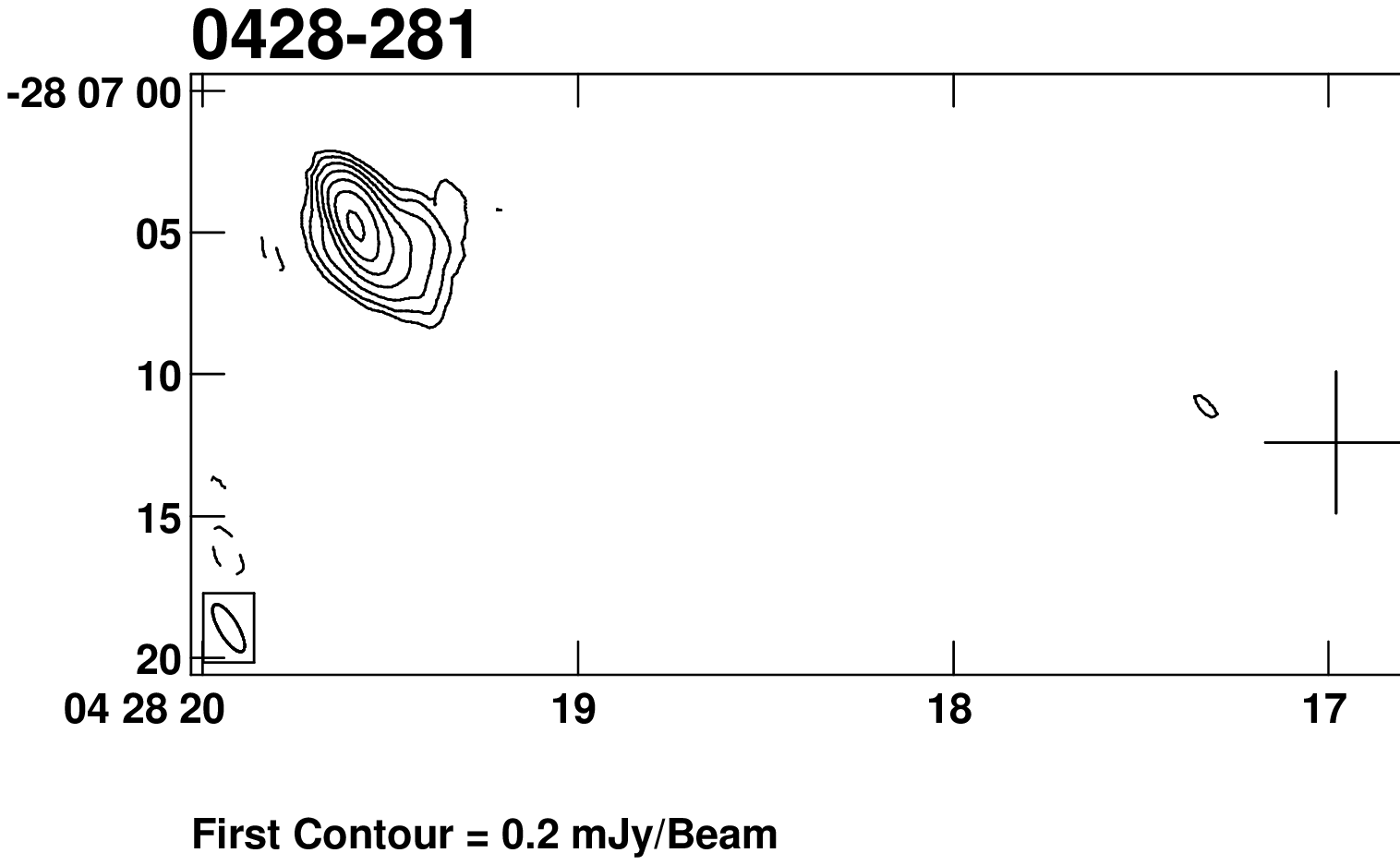,height=5.0in}
\psfig{figure=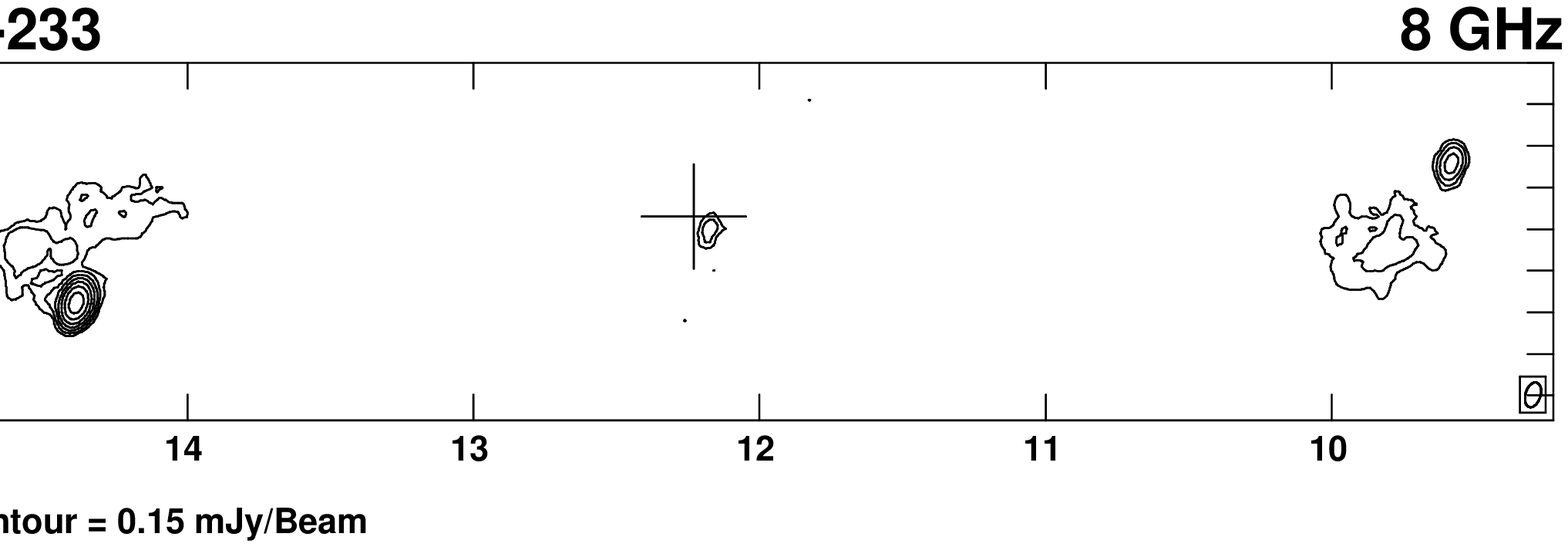,height=5.0in}
\hspace{1.0in}
\psfig{figure=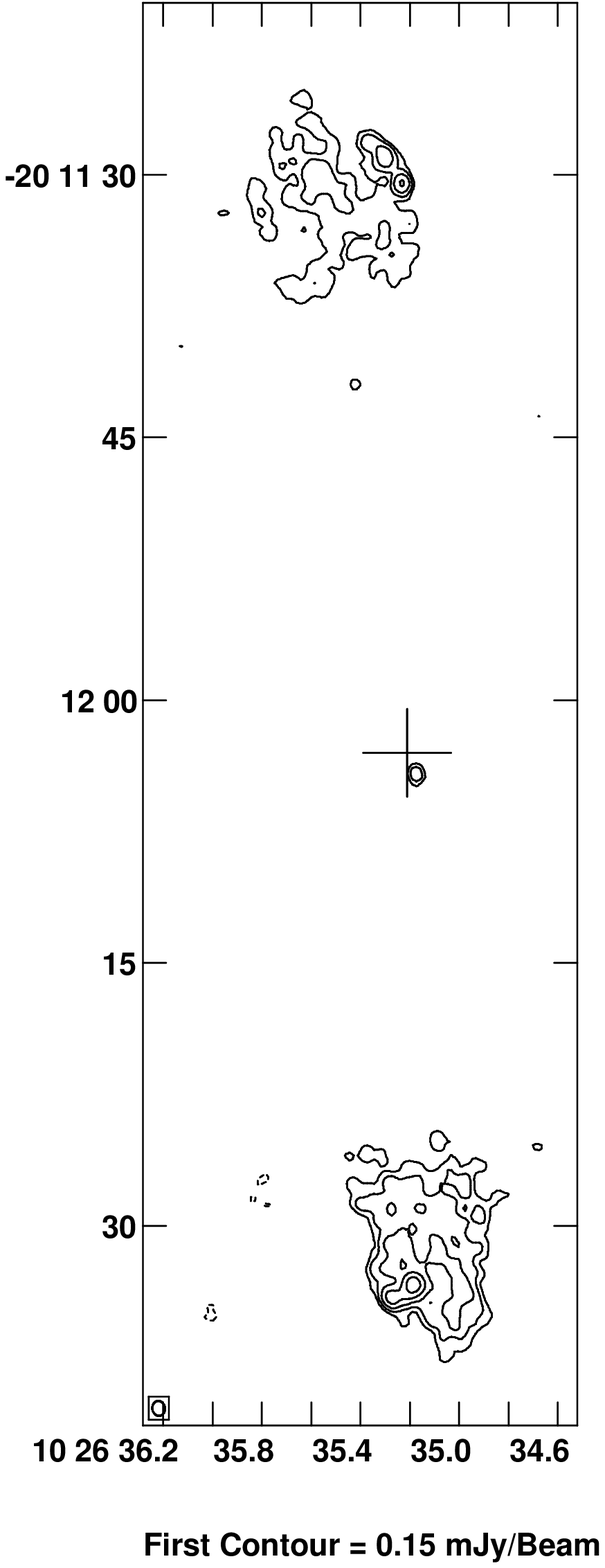,height=4.5in}
}
\vspace{-1.0in}
\hbox{
\hspace{-0.15in}
\psfig{figure=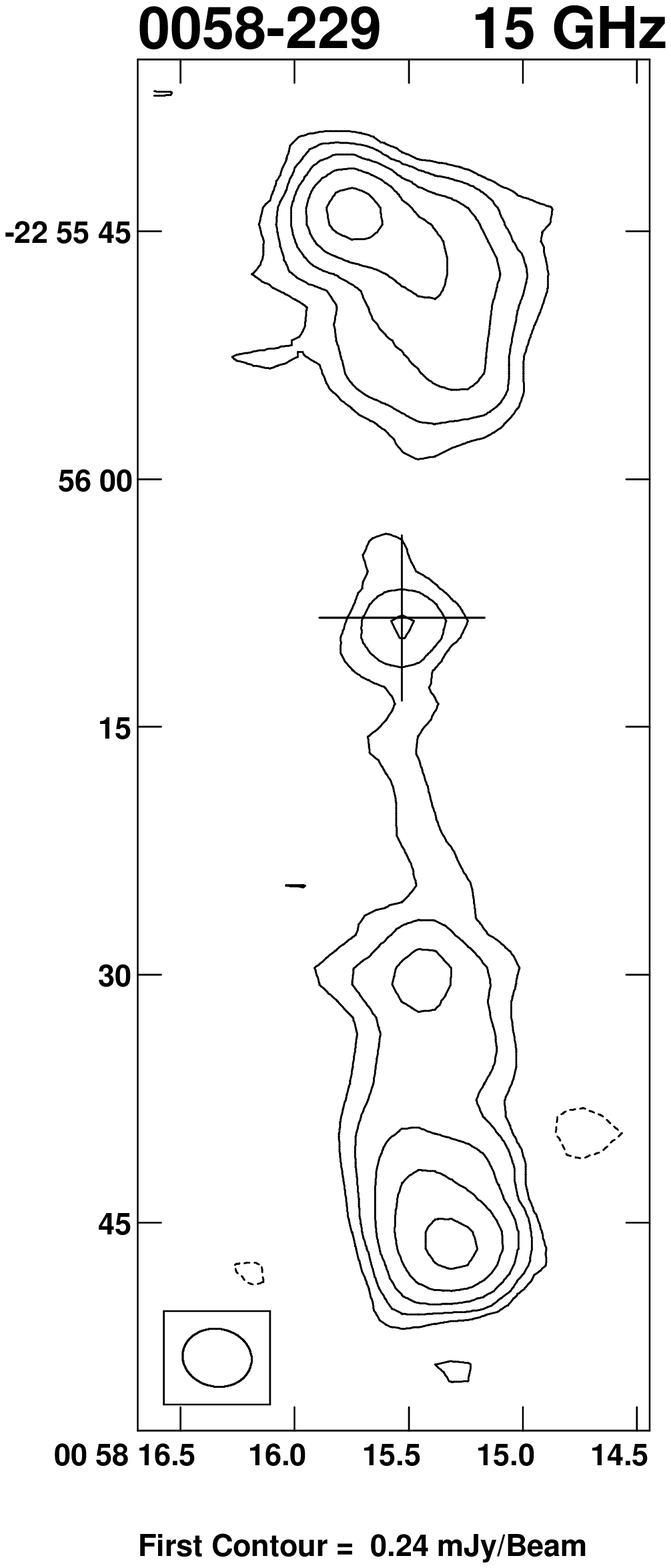,height=3.5in}
\hspace{0.15in}
\psfig{figure=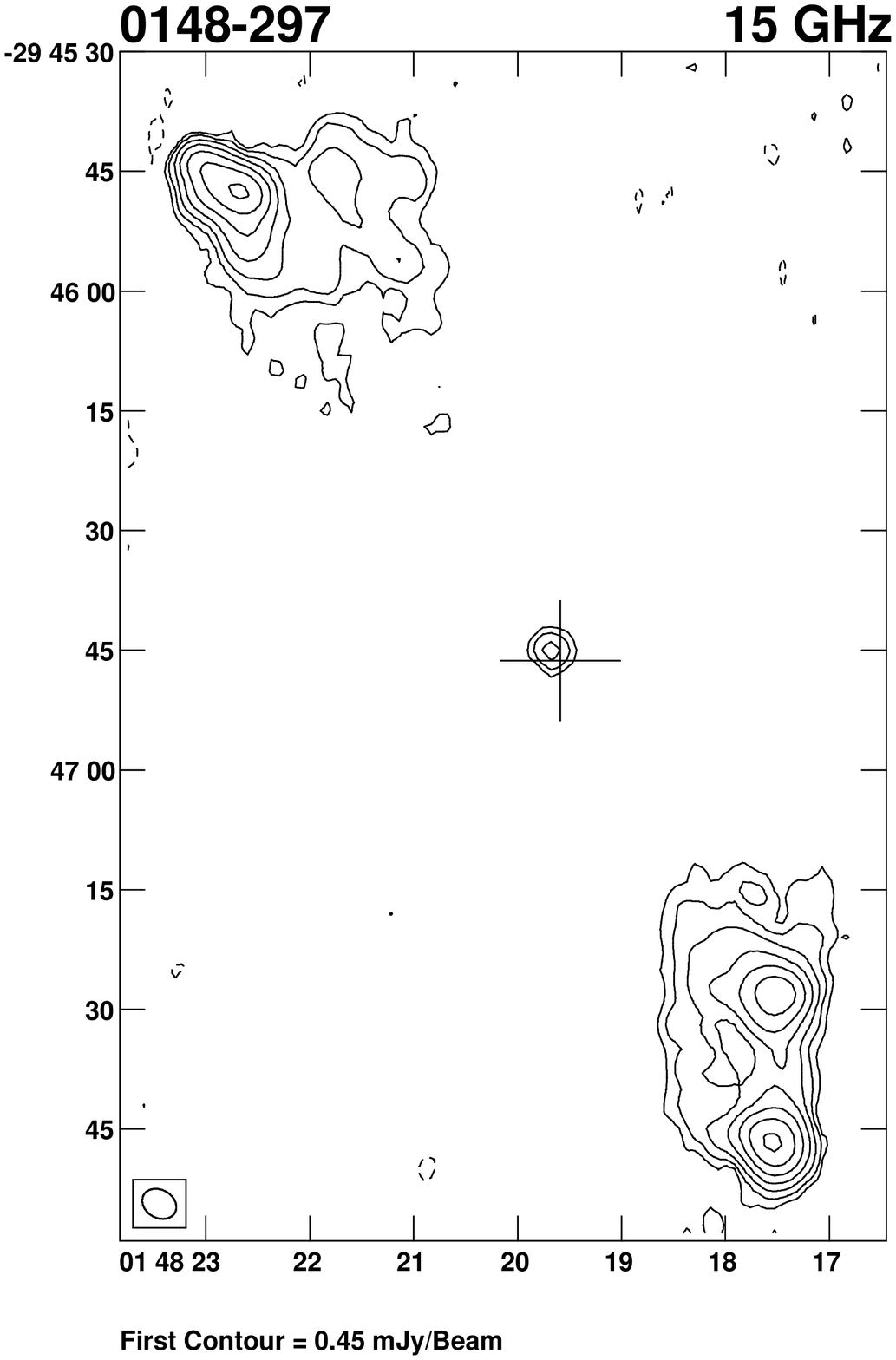,height=3.5in}
\hspace{0.125in}
\psfig{figure=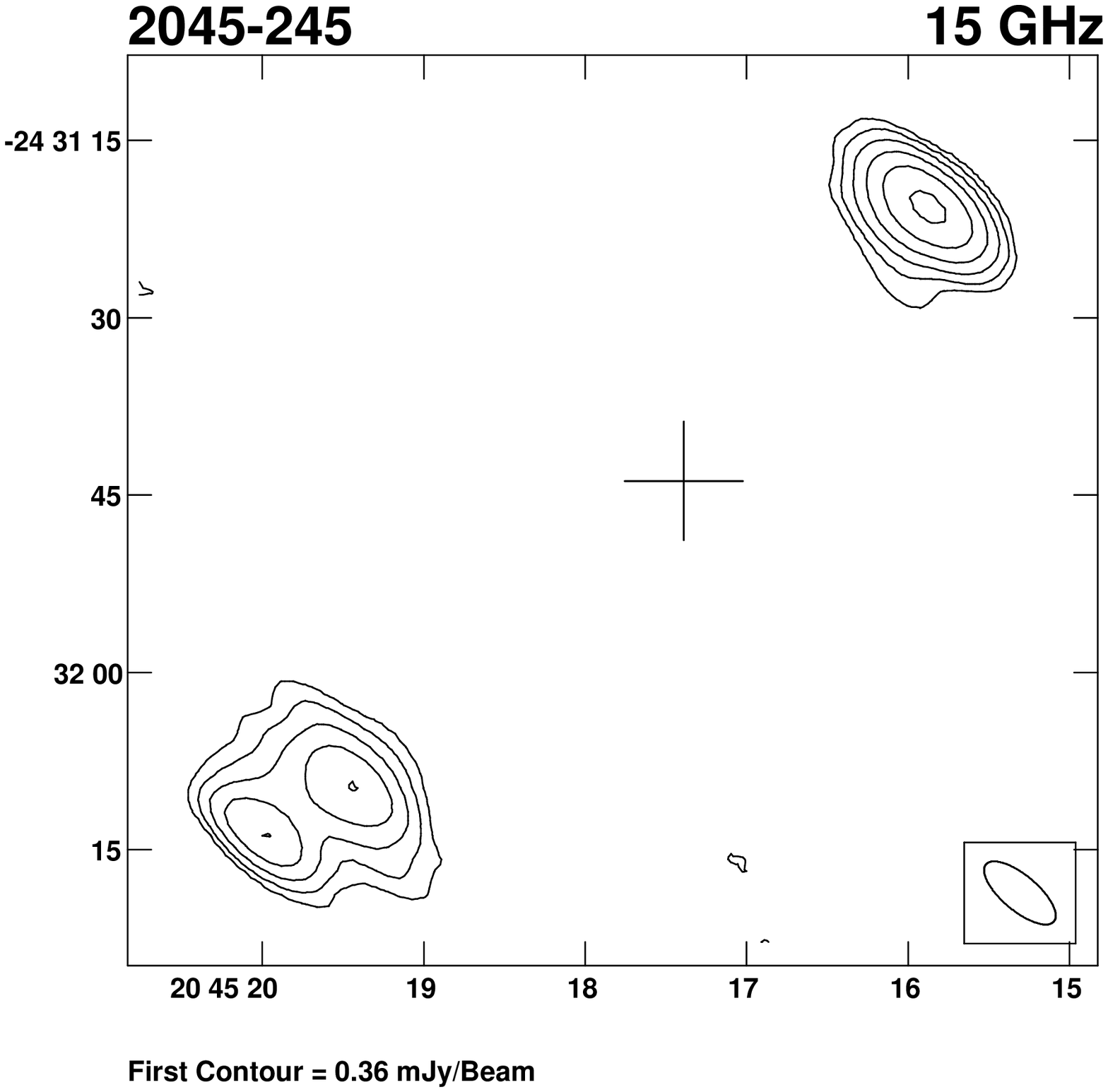,height=3.0in}
}
\caption{Images of the sources observed at X- and U-Band (8 GHz and 15
GHz). The contour levels are -2, -1, 1, 2, 4, 8, 16, 32, 64, 128, 256, 512
mJy/beam times the base level given below each image. The $x$ axis is
right ascension and $y$ axis is declination in B1950 co-ordinates. The
cross denotes the positions of the optical objects. For the 8-GHz image of 
0428$-$281, the cross denotes the position of the optical object given 
in McCarthy et. al 1996 (the galaxy 'A' of Figure 7).} 
\end{figure*}

\begin{table}
\centering
\caption{Observing log: Optical}
\begin{tabular}{llllll}
             &    &      &      &         &              \\
Source       & Id & z    &filter& t       & Date of \\
Name         &    &      &/band & min     & Observation \\
             &    &      &      &         &              \\
0346$-$297   &  G &0.416 & 7052 &  70     & 1997 Mar 7,8 \\
             &    &      & $r$  &  13     & 1997 Mar 8   \\
0428$-$281   &  G &0.65  & 6147 &  75     & 1997 Mar 7   \\
             &    &      & $r$  &  10     & 1997 Mar 7   \\
1023$-$226   &  G &0.586 & 7935 & 140     & 1997 Mar 5   \\
             &    &      & $r$  & 26      & 1997 Mar 5   \\
1052$-$272   &  Q &1.103 & $i$  & 60      & 1997 Mar 8   \\
             &    &      & $V$  & 40      & 1997 Mar 8   \\
1126$-$290   &  G & 0.41 & $i$  & 20      & 1997 Mar 7   \\
             &    &      & $V$  & 25      & 1997 Mar 7   \\
1226$-$297   &  Q & 0.749& 6520 & 57      & 1997 Mar 5   \\
             &    &      & $r$  & 20      & 1997 Mar 5   \\
             &    &      & $V$  & 5       & 1997 Mar 5   \\
1232$-$249   &  Q & 0.352& 6781 & 131     & 1997 Mar 8   \\
             &    &      & $r$  & 20      & 1997 Mar 8   \\
             &    &      & $V$  & 5       & 1997 Mar 8   \\
1247$-$290   &  Q & 0.77 & $i$  & 20      & 1997 Mar 6   \\
             &    &      & $V$  & 40      & 1997 Mar 6   \\
1257$-$230   &  Q & 1.109& $i$  & 40      & 1997 Mar 7   \\
             &    &      & $V$  & 30      & 1997 Mar 7   \\
1358$-$214   &  G & 0.5  & 7531 & 80      & 1997 Mar 4   \\
             &    &      & $i$  & 42      & 1997 Mar 5   \\
             &    &      & $r$  & 42      & 1997 Mar 5   \\
\end{tabular}
\end{table}

\subsection{Optical observations } 
Optical narrow-band imaging at the
redshifted O[II] or O[III] lines with appropriate filters and broad-band
imaging using $r$, $i$ and V filters were done for a subset of sources
from the original sample. For broad-band imaging the $r$ filter of the
Thuan \& Gunn (1976) system, the $i$ filter of Wade et al. (1979) system
and the $V$ filter of the Johnson (1963) system were used. For narrow-band
observations, filters centered at 6147, 6520, 6781, 7052, 7531 and 7935 \AA
\ with widths of about 70 to 80 \AA , \ courtesy of the Lick Observatories, were
used. The observations were carried out with the 2.5m Du Pont Telescope at
Las Campanas Observatory (LCO), Chile. A 2048 $\times$ 2048, thinned and
UV flooded, Textronix CCD with 24\,$\mu$m $\times$ 24\,$\mu$m sized pixels
was used.  The 2048 $\times$ 2048 image was rebinned to derive 1024
$\times$ 1024 images. Astrometry was performed using the Digitized Sky
Survey (DSS) images which have a pixel size of about 1.$^{\prime\prime}$7.  
Using these DSS images a pixel scale of about 0.$^{\prime\prime}$52
pixel$^{-1}$ was derived for our final images. This corresponds to total
field size of 532 $\times $ 532 arcsec$^2$ per CCD frame.  We have checked
our identifications with the finding charts given in McCarthy et al.
(1996) and Kapahi et al. (1998a) for radio galaxies and quasars
respectively.  For CCD image pre-processing, the bias correction was done
using the 16 column overscan region of the individual CCD frames.
Flat-fielding was done for each filter using master dome flats, which were
obtained by averaging corresponding individual dome flats.  Multiple
frames of individual sources were properly aligned and summed to produce
the final image. Absolute photometry was not performed on our images as we
are interested only in the morphology of the source and in the
distribution of the emission-line gas around the source. Instead, each
final image was normalised with respect to its mean value near the source.
The normalised broad-band continuum images were used to subtract from the
normalised narrow-band images to produce the continuum subtracted
narrow-band images, except in one source (1023$-$226) where the narrow-
and broad-band filters did not overlap. The data were analysed using the
{\tt IRAF} package provided by National Optical Astronomy Observatory (NOAO).

The properties of the sources and the observing log are presented in Table
2, which is arranged as follows. Columns 1, 2 and 3: source name, its
optical identification (G for galaxy and Q for quasar) and redshift;
column 4: filters used in the observations; column 5: integration time in
minutes and column 6: date of observation.

\section{Results from the radio observations} 

We present the X- and U-band images for a few sources in Figure 1. 
The images of all the sources are presented in Ishwara-Chandra 1999.
We summarise the restoring beam and the rms noise in each of the images, 
and the total and peak flux densities of the lobes on opposite sides 
of the nucleus in Table 3 for X-band images and in Table 4 for U-band images. 
The sources which have been used in the discussion of Section 3.2 have been
marked with an asterisk.
The core flux densities estimated from these images have been listed 
by Ishwara-Chandra \& Saikia (2000). 

\begin{table}\caption{X band flux densities of lobes}
\begin{tabular}{lrrrrrrr}
Source     & maj   & min   &  PA   &$\sigma$& Cp  & Pk.  &  Tot \\
           &       &       &       &       &     &      &      \\
0325$-$260 & 2.38  & 0.70  & 37    & 43    & N   & 8.8  & 16.0 \\
           &       &       &       &       & S   & 0.6  & 6.4  \\
0346$-$297 & 2.33  & 0.65  & 35    & 95    & N   & $ - $& $-$  \\
           &       &       &       &       & S   & 1.8  & 21.2 \\
0428$-$281$^*$ & 1.90  & 0.66  & 31    & 84    & E   & 15.0 & 39.9 \\
           &       &       &       &       & W   & 24.8 & 64.0 \\
0437$-$244 & 1.89  & 0.58  & 35    & 85    & N   & 2.2  & 19.6 \\
           &       &       &       &       & S   & 5.7  & 19.1 \\
0454$-$220$^*$ & 1.88  & 0.57  & 36    & 311   & N   & 5.2  & 93.0 \\
           &       &       &       &       & S   & 11.2 & 107.0\\
0551$-$226 & 1.76  & 0.56  & 33    & 82    & N   & 0.5  & 2.4  \\
           &       &       &       &       & S   & 0.24 & 1.7  \\
0937$-$250 & 1.20  & 0.72  & $-$21 & 79    & N   & 2.0  & 7.1  \\
           &       &       &       &       & S   & 1.0  & 15.6 \\
0938$-$205 & 1.22  & 0.69  & $-$28 & 78    & N   & 0.3  & 6.1  \\
           &       &       &       &       & S   & 0.8  & 13.7 \\
0947$-$247$^*$ & 1.19  & 0.73  & $-$19 & 138   & N   & 30.3 & 78.6 \\
           &       &       &       &       & S   & 12.7 & 27.8 \\
0955$-$283 & 1.46  & 0.72  & 9     & 86    & E   & 0.8  & 10.1 \\
           &       &       &       &       & W   & 12.2 & 41.4 \\
1022$-$250 & 1.24  & 0.74  & $-$7  & 79    & N   & 0.8  & 6.3  \\
           &       &       &       &       & S   & 0.5  & 5.9  \\
1023$-$226 & 1.30  & 0.68  & $-$31 & 83    & N   & 2.4  & 14.6 \\
           &       &       &       &       & S   & 0.8  & 4.6  \\
1025$-$229 & 0.80  & 0.61  & 53    & 73    & N   & 3.1  & 16.1 \\
           &       &       &       &       & S   & 0.9  & 2.6  \\
1026$-$202 & 0.85  & 0.72  & 9     & 101   & N   & 1.4  & 57.8 \\
           &       &       &       &       & S   & 4.1  & 71.9 \\
1029$-$233$^*$ & 1.22  & 0.72  & $-$17 & 79    & E   & 15.1 & 43.7 \\
           &       &       &       &       & W   & 1.7  & 11.3 \\
1052$-$272$^*$ & 1.12  & 0.71  & 4     & 80    & N   & 1.8  & 33.8 \\
           &       &       &       &       & S   & 1.8  & 31.0 \\
1107$-$218$^*$ & 1.23  & 0.69  & $-$26 & 86    & E   & 17.0 & 23.1 \\
           &       &       &       &       & W   & 9.4  & 11.7 \\
1107$-$227$^*$ & 1.22  & 0.75  & $-$7  & 86    & N   & 17.7 & 32.5 \\
           &       &       &       &       & S   & 8.3  & 14.6 \\
1126$-$290$^*$ & 1.55  & 0.75  & $-$16 & 90    & N   & 5.1  & 17.4 \\
           &       &       &       &       & S   & 14.1 & 51.8 \\
1224$-$208 & 1.31  & 0.67  & $-$32 & 53    & N   & 0.7  & 10.8 \\
           &       &       &       &       & S   & 2.0  & 6.9  \\
1226$-$297 & 1.41  & 0.71  & 6     & 100   & N   & 1.3  & 2.3  \\
           &       &       &       &       & S   & 46.3 & 79.3 \\
1232$-$249$^*$ & 1.21  & 0.7   & 1     & 298   & N   & 33.3 & 204.9\\
           &       &       &       &       & S   & 52.4 & 179.2\\
1247$-$290$^*$ & 1.17  & 0.7   & 6     & 117   & N   & 3.5  & 51.4 \\
           &       &       &       &       & S   & 12.0 & 33.6 \\
1257$-$230$^*$ & 1.16  & 0.73  & 3     & 113   & N   &  2.7 & 22.0 \\
           &       &       &       &       & S   & 45.7 & 91.3 \\
1358$-$214 & 1.34  & 0.73  & 25    & 89    & N   & 2.6  & 15.1 \\
           &       &       &       &       & S   & 2.4  & 15.9 \\
\end{tabular}
\end{table}

\begin{table}\caption{U band flux densities of lobes}
\begin{tabular}{lrrrrrrr}
Source     & maj   & min   & PA    &$\sigma$& Cp & Pk.  &  Tot \\
           &       &       &       &       &     &      &      \\
0017$-$207$^*$ & 4.5   & 3.5   & +70   & 162   & N   & 14.0 & 22.0 \\
           &       &       &       &       & S   & 10.6 & 21.2 \\
0058$-$229$^*$ & 4.2   & 3.5   & +80   & 134   & N   & 5.3  & 17.1 \\
           &       &       &       &       & S   & 4.6  & 16.8 \\
0148$-$297$^*$ & 4.5   & 3.5   & +60   & 223   & N   & 31.2 & 106.5\\
           &       &       &       &       & S   & 32.4 & 148.4\\
0428$-$281$^*$ & 6.0   & 3.5   & $-$70 & 202   & E   & 16.7 & 27.3 \\
           &       &       &       &       & W   & 26.1 & 36.7 \\
0437$-$244$^*$ & 9.0   & 5.0   & $-$25 & 183   & N   & 6.9  & 10.8 \\
           &       &       &       &       & S   & 3.8  & 6.0  \\
0454$-$220$^*$ & 6.0   & 3.0   & $-$70 & 439   & N   & 19.3 & 50.9 \\
           &       &       &       &       & S   & 32.0 & 63.5 \\
0947$-$247$^*$ & 8.0   & 4.0   & $-$50 & 327   & N   & 41.6 & 57.3 \\
           &       &       &       &       & S   & 20.2 & 28.6 \\
1025$-$229 & 8.0   & 4.5   & $-$50 & 272   & N   & 3.9  & 4.8  \\
           &       &       &       &       & S   & 4.0  & 15.3 \\
1026$-$202$^*$ & 8.0   & 4.2   & $-$50 & 214   & N   & 6.9  & 20.4 \\
           &       &       &       &       & S   & 17.4 & 40.3 \\
1052$-$272$^*$ & 6.0   & 5.0   & $-$50 & 229   & N   & 9.0  & 13.8 \\
           &       &       &       &       & S   & 7.9  & 19.8 \\
1107$-$227$^*$ & 6.5   & 4.5   & $-$50 & 200   & N   & 11.3 & 16.2 \\
           &       &       &       &       & S   & 4.8  & 7.2  \\
1126$-$290$^*$ & 6.0   & 5.0   & $-$50 & 230   & N   & 7.4  & 26.9 \\
           &       &       &       &       & S   & 17.4 & 67.6 \\
1226$-$297 & 7.0   & 4.5   & $-$50 & 198   & N   & 1.2  & 1.4  \\
           &       &       &       &       & S   & 40.6 & 47.2 \\
1232$-$249$^*$ & 7.0   & 4.5   & $-$50 & 335   & N   & 50.1 & 112.9\\
           &       &       &       &       & S   & 77.3 & 135.1\\
1247$-$290$^*$ & 7.0   & 5.5   & $-$50 & 288   & N   & 14.5 & 28.0 \\
           &       &       &       &       & S   & 9.0  & 17.1 \\
1257$-$230$^*$ & 7.0   & 4.5   & $-$50 & 282   & N   & 8.8  & 11.2 \\
           &       &       &       &       & S   & 57.5 & 70.1 \\
2035$-$203 & 7.0   & 3.0   & +50   & 279   & E   & 71.2 & 83.0 \\
           &       &       &       &       & W   & 1.4  & 7.3  \\
2040$-$236 & 6.4   & 3.0   & +50   & 274   & E   & 10.7 & 21.9 \\
           &       &       &       &       & W   & 1.0  & 2.3  \\
2045$-$245$^*$ & 7.5   & 3.0   & +50   & 167   & N   & 12.9 & 22.1 \\
           &       &       &       &       & S   & 5.7  & 20.2 \\
2118$-$266 & 7.5   & 3.0   & +50   & 242   & E   & 9.5  & 19.4 \\
           &       &       &       &       & W   & 0.7  & 1.6  \\
2213$-$283$^*$ & 6.5   & 3.0   & +50   & 213   & E   & 9.1  & 33.0 \\
           &       &       &       &       & W   & 8.3  & 14.1 \\
2311$-$222$^*$ & 5.0   & 3.0   & +60   & 237   & E   & 68.9 & 85.2 \\
           &       &       &       &       & W   & 8.4  & 13.7 \\
2325$-$213$^*$ & 5.0   & 3.0   & +60   & 202   & N   & 8.8  & 21.6 \\
           &       &       &       &       & S   & 13.2 & 34.6 \\
2338$-$290 & 5.0   & 3.5   & +60   & 290   & N   & 6.7  & 12.1 \\
           &       &       &       &       & S   & 1.5  & 5.3  \\
\end{tabular}
\end{table}

\subsection{The Liu-Pooley effect} 

\begin{figure}
\psfig{figure=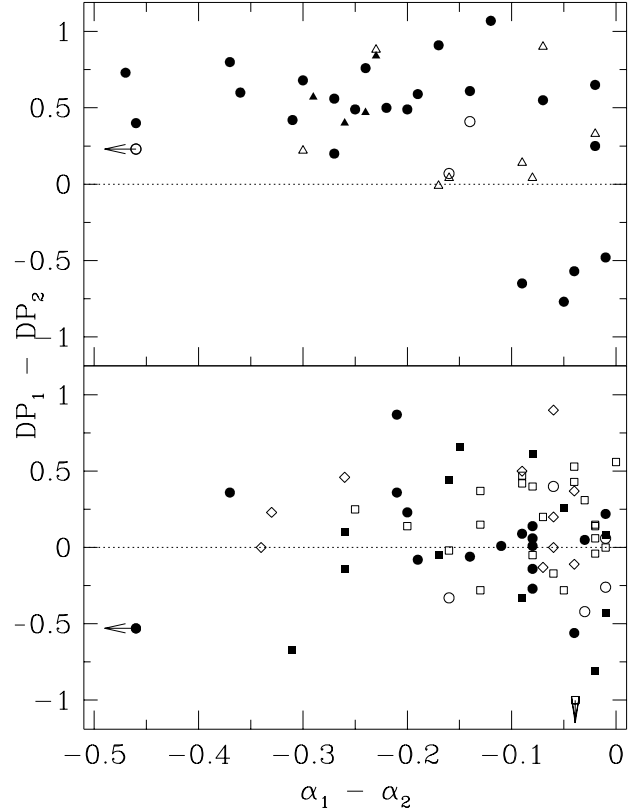,angle=0,height=4.50in} 

\caption{The Liu-Pooley relationship for the samples of small (upper
panel) and large (lower panel) sources. The triangles, circles, diamonds
and squares represent the sources from the Liu-Pooley, Garrington et al.,
Pedelty et al. and the present MRC samples respectively. The open and
filled symbols denote galaxies and quasars respectively.} 
\end{figure}

We have investigated the trend for the lobe with the flatter spectral
index to be less depolarized, also known as the Liu-Pooley effect for our
sample of sources, along with those from Liu \& Pooley (1991), Garrington
et al. (1991) and Pedelty et al. (1991). The sources in these samples have
all been observed with the VLA between $\lambda$20 and 6 cm and the
authors have tabulated the DP or depolarization values and spectral
indices for the individual lobes. DP is defined as $m_{20}$/$m_{6}$, where
$m$ is the degree of polarization at the two wavelengths. The results from
the different samples are summarized in Table 5, which is arranged as
follows. Column 1: the different samples where LP91 denotes Liu \& Pooley,
G91-S and G91-L denote the samples of small and large sources from
Garrington et al. (1991), P91 the sources from Pedelty et al. (1991) and
MRC the present sample; column 2: the total number of galaxies and quasars
in each of these samples; column 3: the number of galaxies and quasars
which show the Liu-Pooley effect. The typical error in the difference in
DP and spectral index is about 0.1.  The first number indicates the number
of sources where the differences in DP and spectral index are larger than
0.1, while the number in brackets indicates the total number of sources
consistent with the the Liu-Pooley relation.  Column 4: the median values
of $\Delta\alpha$, the difference in spectral indices between the lobes,
for the galaxies and quasars; columns 5 to 8: the corresponding median
values of $\Delta$DP defined to be the difference in depolarization
between the two lobes, projected linear size in kpc, redshift and the
fraction of emission from the core at an emitted frequency of 8 GHz.

We first consider the samples consisting of small sources, namely from the LP91
and G91-S samples (Figure 2, upper panel) where the median linear sizes are
about 85 and 88 kpc respectively. 
Among the sources whose values of $\Delta$DP and
$\Delta\alpha$ are $>$0.1, all the 4 galaxies and the 20 quasars are
consistent with the Liu-Pooley effect. If we consider all the sources,
about 91 per cent of the galaxies and 85 per cent of the quasars are
consistent with the Liu-Pooley effect. In the samples consisting of larger
sources (Figure 2, lower panel), namely the G91-L, P91 and the IC98 samples, 
where the median linear sizes are about 347, 350 and 595 kpc respectively,
we find that the
effect is weaker, the difference in DP also being smaller (see Figure 3).
Among those whose values of $\Delta$DP and
$\Delta\alpha$ are $>$0.1, 6 of the 8 galaxies (75 per cent) and 7 of the
10 quasars show the Liu-Pooley effect. The corresponding values for the
entire data are 58 and 59 per cent for the galaxies and quasars
respectively.

Ideally, since the DP values depend on wavelength, and the objects are at
different redshifts one should estimate the DP values at fixed wavelengths
in the emitted frame. However, the $m - \lambda$ curves for the individual
lobes are usually not known. We have therefore examined the trends for
objects in a narrower range of redshift, namely 0.4 to 1, and
find the results to be similar to that of the entire sample where the 
redshift ranges from 0.06 to 2.3.

\begin{figure}
\psfig{figure=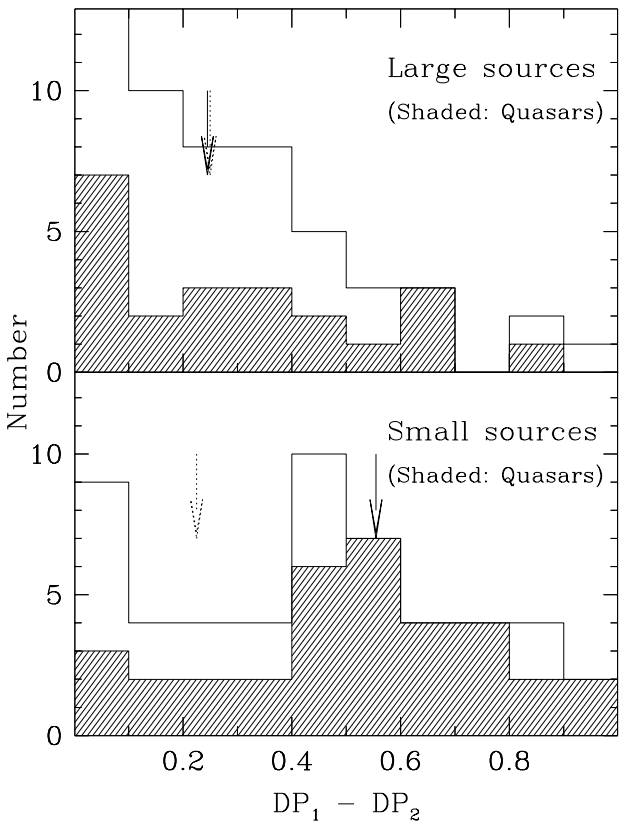,height=4.5in}
\caption{Distributions of the absolute difference in DP between the 
two lobes for the large and small sources. The dotted and continuous 
arrows mark the median values for the galaxies and quasars respectively.}
\end{figure}

\begin{figure}
\vbox
{
\psfig{figure=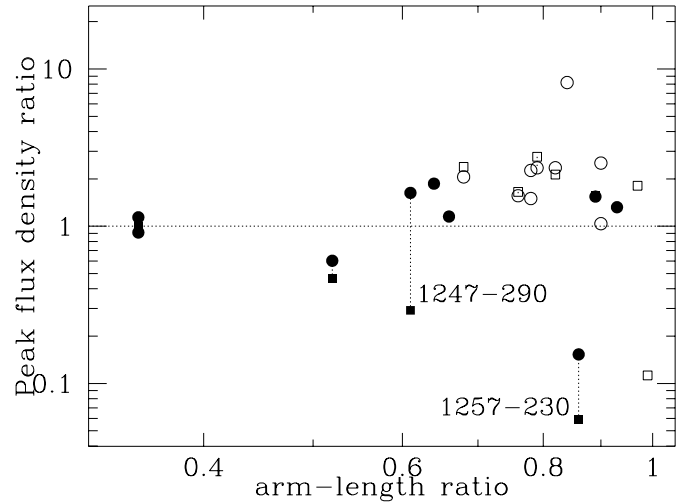,angle=270,height=2.75in}
}
\caption{Ratio of peak brightness at X- and U-bands of the nearer 
lobe to the farther one against the corresponding arm-length ratio. 
The open and filled circles denote galaxies and quasars from U 
band observations, whereas open and filled squares denote galaxies 
and quasars from X band observations.}
\end{figure}

\begin{table*} \caption{Parameters for all Samples}
\begin{tabular}{|l| rl| rl| rr| rr| rr| rr| rr| rr|}
\multicolumn{1}{c}{Samp}& \multicolumn{2}{c}{Number} &\multicolumn{2}{c}{L$-$P} & \multicolumn{2}{c}{$\alpha_1 - \alpha_2$} & \multicolumn{2}{c}{$\mid$DP$_1$ - DP$_2$$\mid$} & \multicolumn{2}{c}{Linear size} & \multicolumn{2}{c}{redshift} & \multicolumn{2}{c}{f$_c$} \\
\hline
      & G & Q & G & Q  &  G      &  Q     & G   &  Q   &    G & Q   &  G   & Q    &  G     &   Q    \\
      &   &   &   &    &         &        &     &      &      &     &      &      &        &        \\
\multicolumn{1}{c}{(1)}& \multicolumn{2}{c}{(2)} &\multicolumn{2}{c}{(3)} & \multicolumn{2}{c}{(4)}
  & \multicolumn{2}{c}{(5)} & \multicolumn{2}{c}{(6)} & \multicolumn{2}{c}{(7)} & \multicolumn{2}{c}{(8)} \\
\hline
LP91 & 9 & 4 & 3(8) & 4(4)  & $-$0.16 & $-$0.25& 0.22& 0.52 &  85  & 94  & 0.9  & 1.05 & 0.00339& 0.0978  \\
      &   &   &   &    &         &        &     &      &      &     &      &      &        &         \\
G91$-$S& 2 & 23& 1(2) & 16(19) & $-$0.15 & $-$0.20& 0.24& 0.59 &  148 & 87  & 0.8  & 1.618& 0.0034 & 0.1198 \\
      &   &   &   &    &         &        &     &      &      &     &      &      &        &         \\
G91$-$L&5  & 17& 0(2) & 4(7)  & $-$0.03 & $-$0.09& 0.33& 0.14 & 264  & 373 & 0.689& 0.743& 0.0032 & 0.2309  \\
      &   &   &   &    &         &        &     &      &      &     &      &      &        &         \\
P91   & 10&   & 2(6) &    & $-$0.07 &        & 0.22&      &  330 &     & 1.026&      & 0.0019 &         \\
      &   &   &   &    &         &        &     &      &      &     &      &      &        &         \\
MRC   &23 & 12& 4(14) & 2(6)  & $-$0.06 & $-$0.12& 0.25& 0.38 &595   & 583 & 0.73 & 0.738& 0.0012 & 0.073   \\
\hline
\end{tabular}
\end{table*}

Considering all the sources which are $<$300 kpc, we have examined whether
the Liu-Pooley effect depends on the redshift of the objects. 19 of the 26
objects with redshift $<$1.15, the median value for these sources, show
the effect compared to 19 of the 25 with larger redshifts. Confining
ourselves to only those whose values of $\Delta$DP and $\Delta\alpha$ are
$>$0.1 shows that 10 of 11 objects with redshift $<$1.15, and 15 of 16
objects with larger redshifts show the effect. There appears to be no
dependence on redshift.

A possible interpretation of the Liu-Pooley effect has been in terms of
relativistic motion of the hotspots, so that the one which is on the
approaching side is less depolarized and also has a flatter spectral
index. One might be tempted to infer that the tendency for the Liu-Pooley
effect to be stronger for the smaller sources is consistent with this
interpretation since the small sources might be expected to be inclined at
smaller angles to the line of sight. However, we note that considering the
samples consisting largely of smaller and larger sources separately, there
is no significant difference in the number of objects which show the
Liu-Pooley effect between galaxies and quasars. We have also examined
whether the Liu-Pooley effect depends on the fraction of emission from the
core, which is being used as a statistical measure of the orientation of
the source axes to the line of sight, but find no significant dependence
on it.

The lack of a difference in the strength of the correlation between
galaxies and quasars suggests that Doppler effects may not be the only
factor. It is relevant to note from Table 5, that the absolute values of
$\Delta$DP and $\Delta\alpha$ do tend to be larger for quasars compared to
galaxies of similar size and redshift, suggesting that Doppler effects do
play a role in the location of points in the Liu-Pooley diagram. The
tendency for the jet side lobe to be flatter as well as less depolarized
suggests that Doppler effects could provide a viable explanation of the
Liu-Pooley relation for quasars. At present, it seems necessary to
postulate intrinsic differences in the oppositely-directed lobes of radio
galaxies which affect their spectra as well as depolarization properties.

\begin{figure*}
\vbox{
\hbox{
\psfig{figure=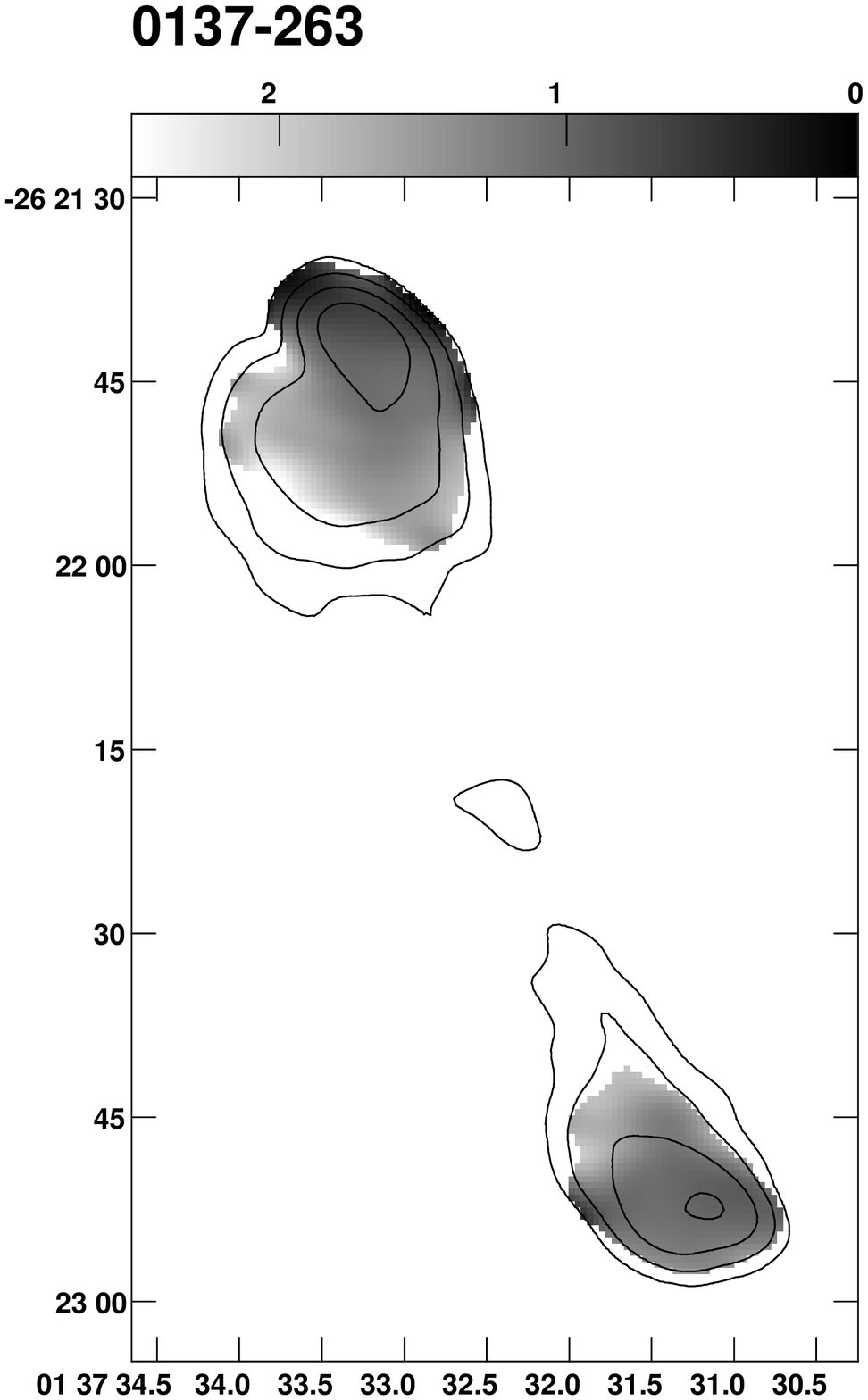,height=2.50in}
\hspace{0.5in}
\psfig{figure=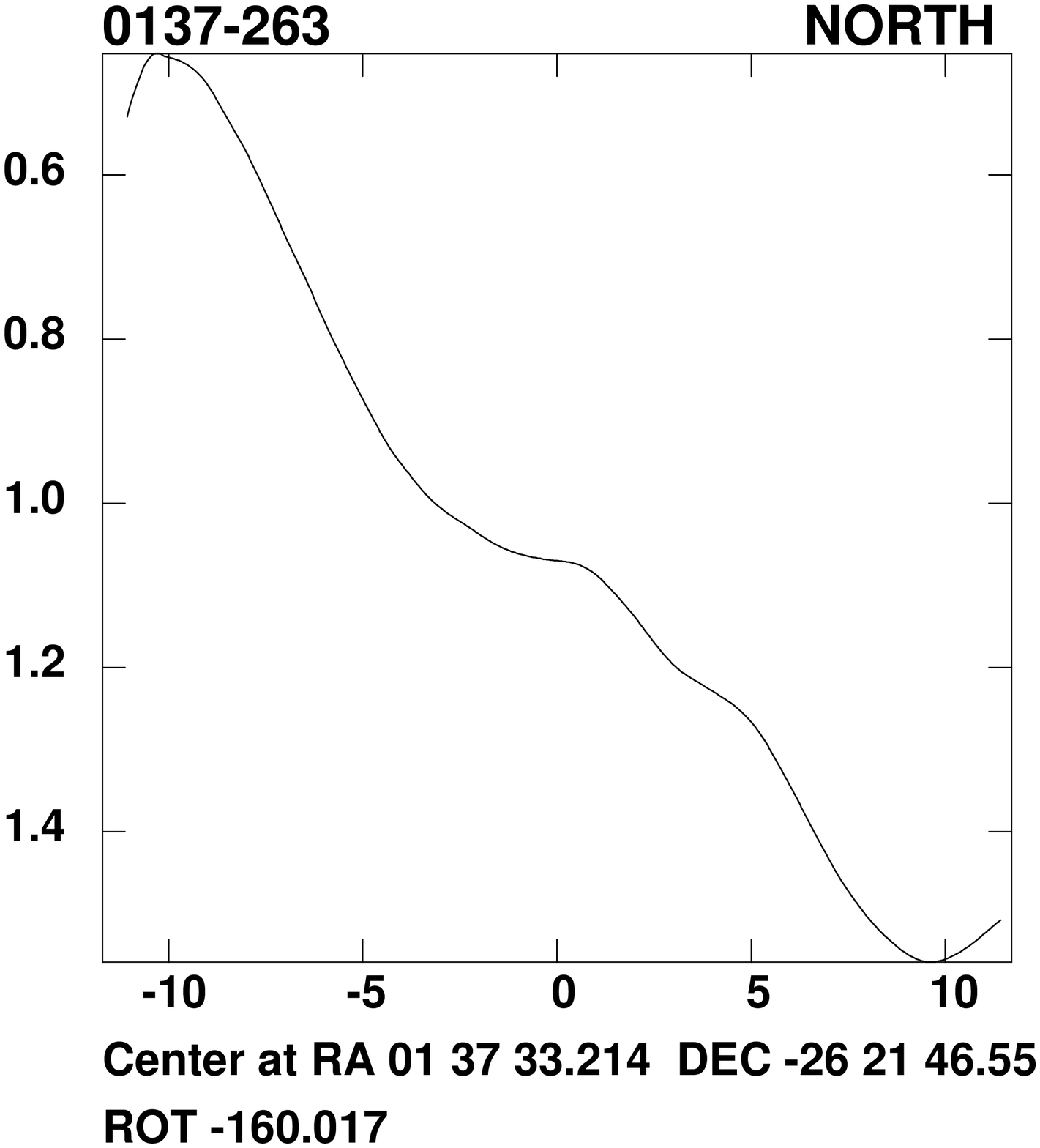,height=2.00in}
\hspace{0.5in}
\psfig{figure=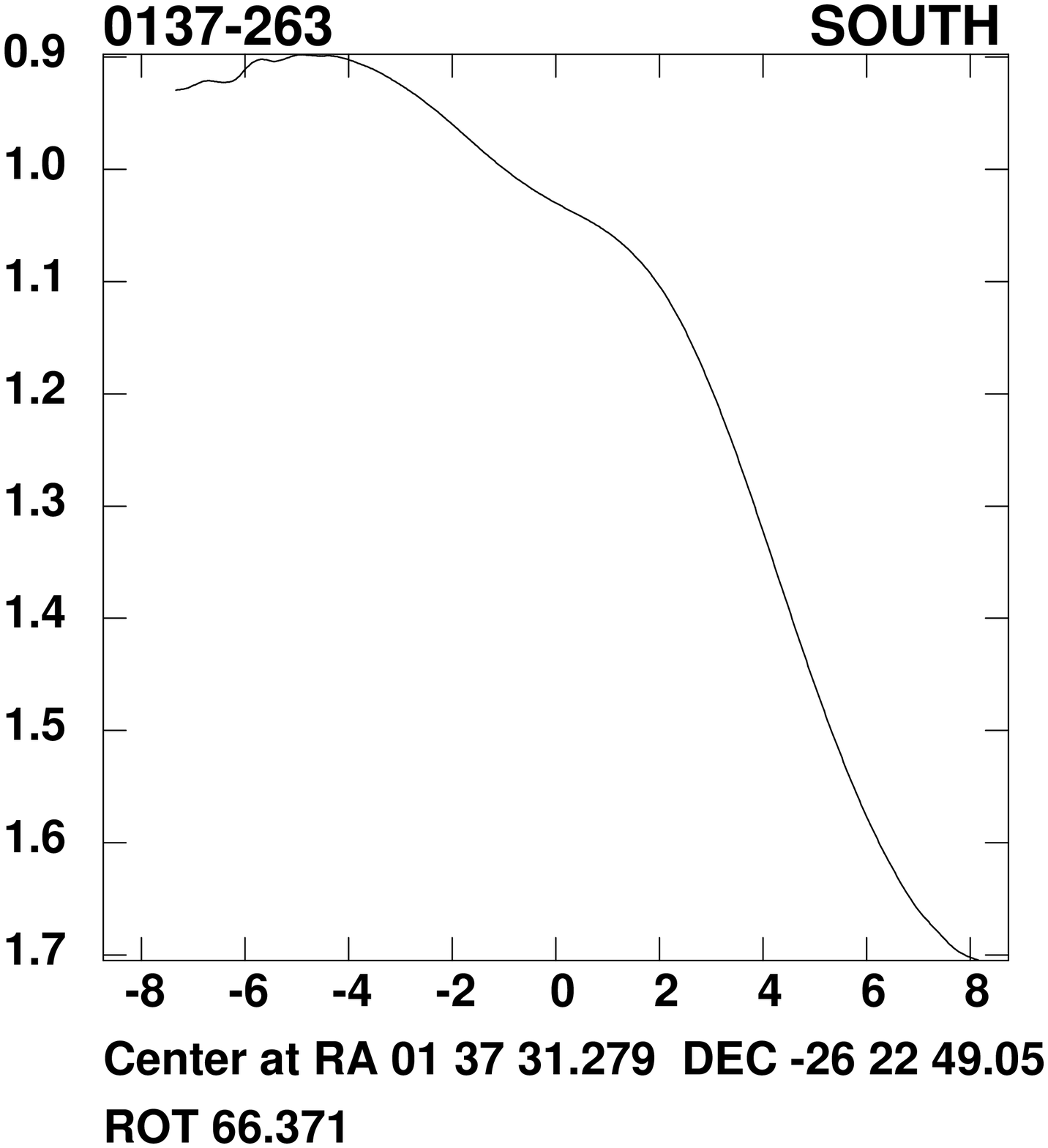,height=2.00in}
}
\hbox{
\psfig{figure=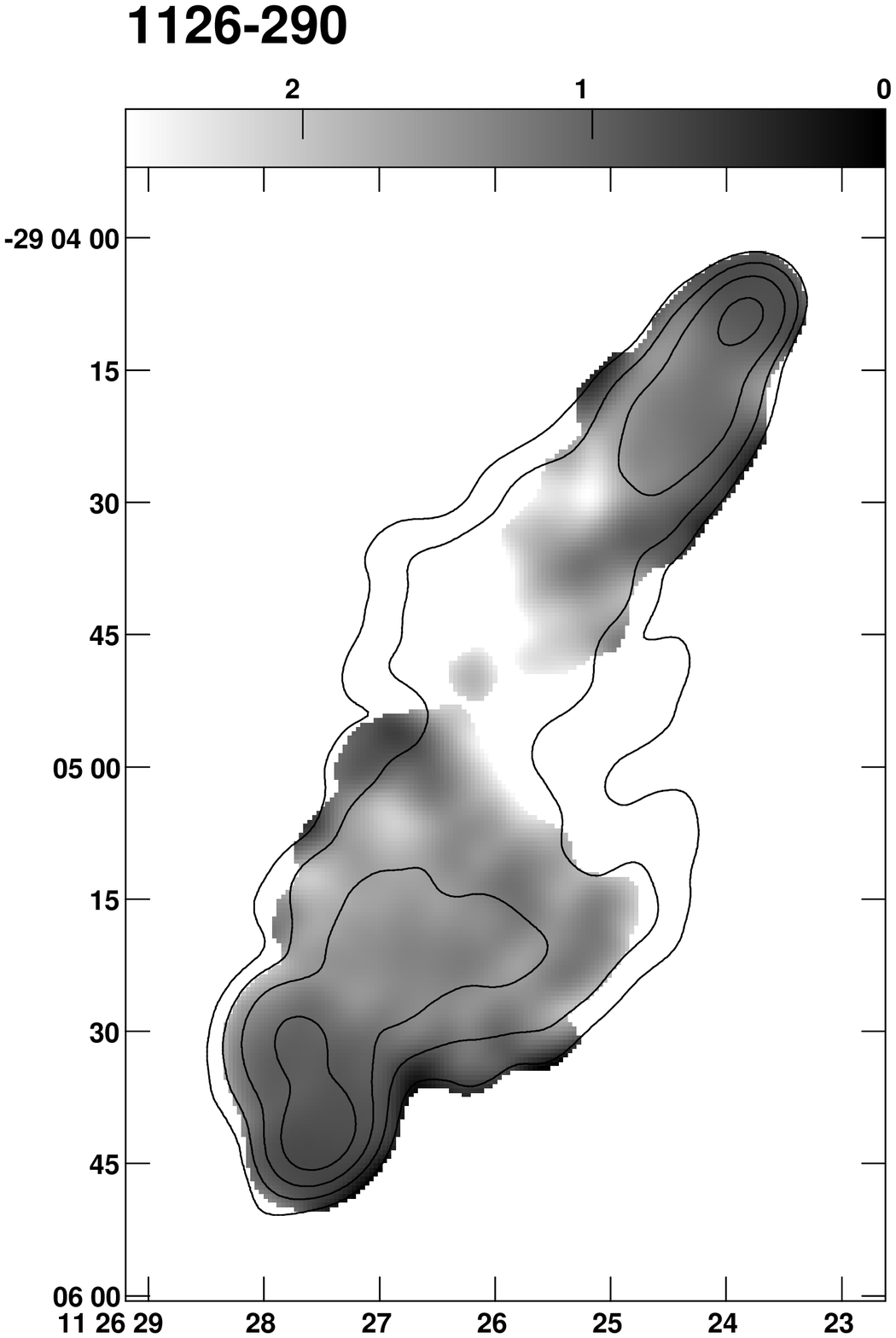,height=2.50in}
\hspace{0.5in}
\psfig{figure=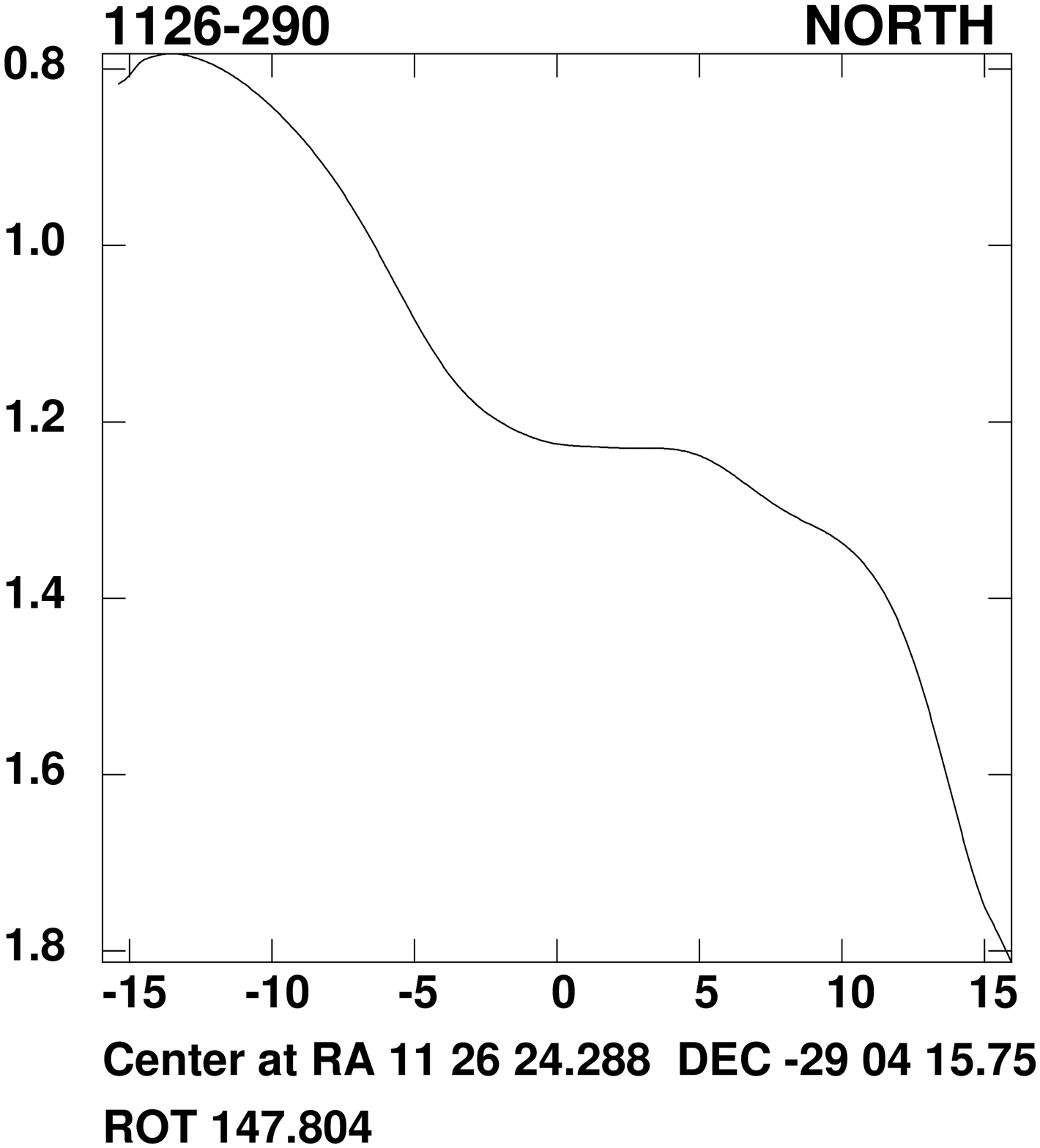,height=2.00in}
\hspace{0.5in}
\psfig{figure=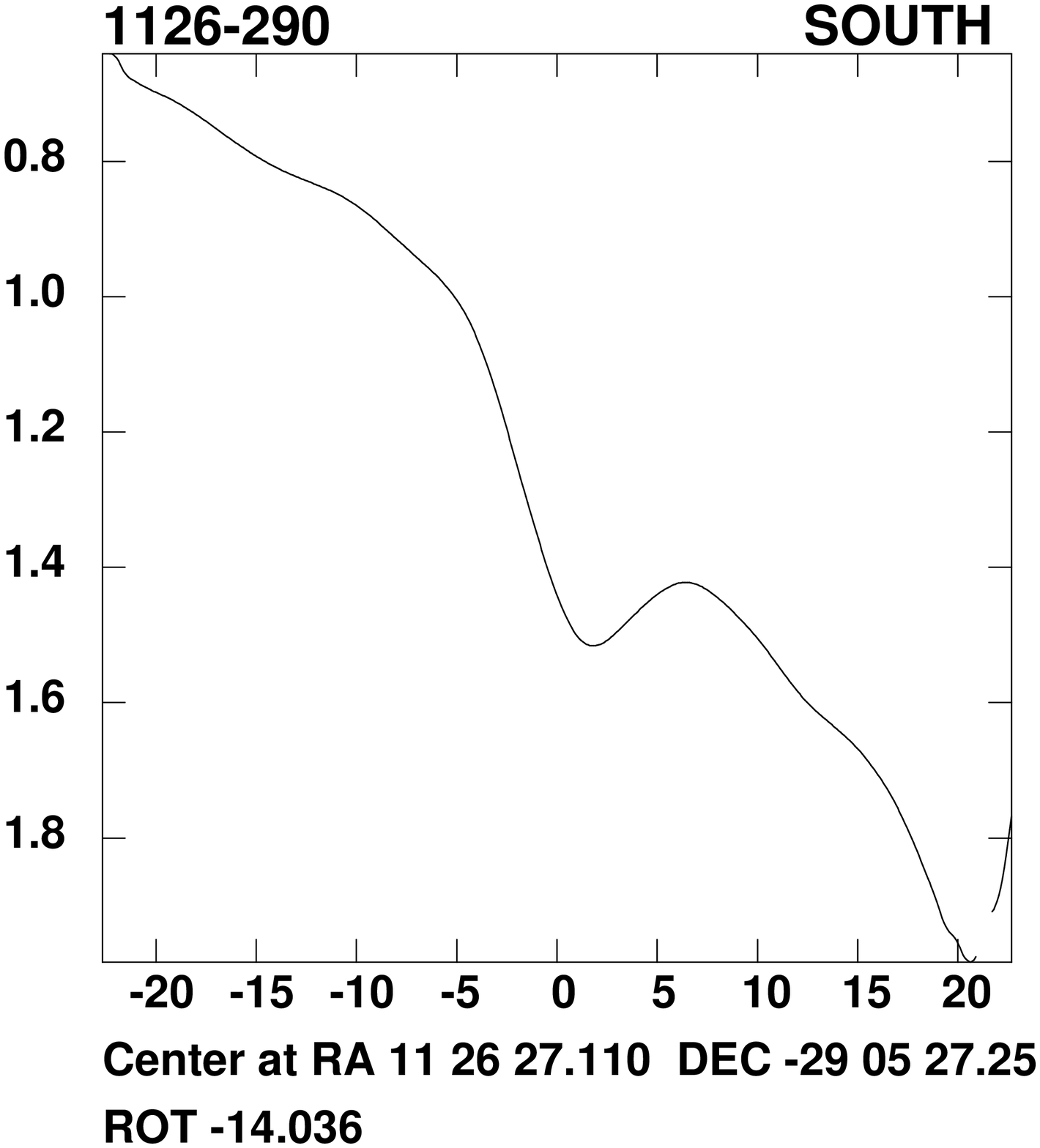,height=2.00in}
}
}
\caption{Two typical examples of spectral index gradients seen in our
sample of sources. The spectral index values are shown in grey
superimposed on the $\lambda$20 cm contour images. The slices illustrate
the variation in spectral indices along the lobes.}
\end{figure*}

\subsection{Brightness and arm-length asymmetry} 

From VLA observations at
L- and C-bands with an angular resolution of about 5 arcsec, IC98 reported
a trend for the nearer component to be brighter and more depolarized. This
is consistent with the idea that these sources are evolving in an
asymmetric environment leading to a higher luminosity for the nearer lobe
which is possibly traversing through the denser medium. In Figure 4 we
show the relationship for the sources observed in the U-band with an
angular resolution of about 5 arcsec, and for the sources observed in the
X-band with a higher resolution of about 1 arcsec. To minimise the errors,
we have considered only the strong hotspots in both bands with a peak flux
density greater than about 10 times the lowest contour level (sources marked
with an asterisk in the Table 3 and 4). The minimum
values of the hotspot flux density in the U- and X-bands are approximately
5 and 1.5 mJy respectively. The flux-density ratio - armlength ratio
diagram for the U-band sources shows a similar relationship reported
earlier (Fig 6 of IC98), although the number of sources is now much
smaller. The trend is stronger for galaxies than the quasars, where the
effects of relativistic beaming is more significant.
The brighter component is nearer in all the 9 galaxies and 6 of the 9
quasars. We have checked these trends at a fixed frequency in the emitted 
frame using the estimates of the hotspot spectral indices from our L- and 
C-band observations, and find that the results are almost identical.

However, in the higher-resolution X-band observations, one of the five
quasars and five of the six radio galaxies have the brighter component
closer to the nucleus. At higher resolutions, where the emission
is dominated by the hotspots, the effects of mild relativistic motion of the
hotspots appear to be more noticeable in quasars. 
Two quasars move significantly along the y-axis, and are shown in 
the Figure 4. The numbers are small and need to be investigated 
more carefully using larger samples. 

\subsection{Spectral ageing}

The spectral age, which is estimated from the change in the spectral index
due to radiative and other losses, is the time elapsed since the electrons
were last accelerated. The electrons are believed to be accelerated at the
hotspot and `age' as they diffuse outwards. These electrons form the
backflow from the hotspot towards the nucleus leading to the formation of
lobes and bridges of radio emission. From a steepening of the radio
spectrum due to radiative losses along this backflow, it is possible, in
principle, to estimate the spectral age.  However, there are several
caveats in interpreting these spectral ages as the dynamical ages of these
sources (cf. Blundell \& Rawlings 2000, and references therein).
Nevertheless, we have estimated the spectral ages and the relative
velocity of advance of the hotspots, usually in a region which is within a
few beamwidths of the peak of emission in the hotspots. We derive the
spectral ages for our sample of sources using the formalism of Myers \&
Spangler (1985), assuming initialy that the change in spectra are only due
to synchrotron losses. The spectral index maps and corresponding spectral
index slices for two representative sources are presented in Figure 5,
demonstrating the range in the spectral index gradients.

\begin{table}
\caption{Age, magnetic field and velocity estimates }
\begin{tabular}{l l l r l r l }
Source &  z      & B$_{eq}$ & Age   & Vel    & Age$^*$ & Vel$^*$ \\
Name   &         &  nT      & Myr   & 0.1c   &  Myr    &  0.1c   \\
       &         &          &       &        &         &         \\
0017$-$207 N  &  0.545  &  0.762   &  14.0 & 0.18  &    7.0  &  0.36 \\
0017$-$207 S  &  0.545  &  0.709   &  47.5 & 0.07  &   22.0  &  0.15 \\
0058$-$229 N  &  0.706  &  0.718   &  32.5 & 0.14  &   12.1  &  0.36 \\
0058$-$229 S  &  0.706  &  0.633   &  50.0 & 0.20  &   15.8  &  0.64 \\
0133$-$266 N  &  1.530  &  1.326   &  10.5 & 0.36  &    3.1  &  1.22 \\
0133$-$266 S  &  1.530  &  1.254   &   9.5 & 0.54  &    2.6  &  1.99 \\
0137$-$263 N  &  1.100  &  1.106   &  24.0 & 0.35  &    9.1  &  0.91 \\
0137$-$263 S  &  1.100  &  0.977   &  20.0 & 0.27  &    6.5  &  0.84 \\
0148$-$297 N  &  0.410  &  0.738   &  34.5 & 0.17  &   19.8  &  0.30 \\
0148$-$297 S  &  0.410  &  0.734   &  22.5 & 0.38  &   12.9  &  0.67 \\
       &         &          &       &        &         &         \\
0325$-$260 N  &  0.638  &  0.708   &  22.0 & 0.35  &    8.9  &  0.85 \\
0325$-$260 S  &  0.638  &  0.644   &  28.5 & 0.29  &   10.3  &  0.81 \\
0346$-$297 S  &  0.413  &  0.682   &  38.0 & 0.25  &   20.2  &  0.48 \\
0428$-$281 E  &  0.650  &  1.062   &  22.5 & 0.27  &   13.5  &  0.46 \\
0428$-$281 W  &  0.650  &  1.174   &  22.5 & 0.42  &   14.5  &  0.65 \\
0437$-$244 N  &  0.840  &  0.737   &  31.5 & 0.20  &   10.0  &  0.65 \\
0437$-$244 S  &  0.840  &  0.608   &  14.3 & 0.17  &    3.4  &  0.72 \\
0454$-$220 S  &  0.533  &  1.269   &  18.5 & 0.35  &   13.7  &  0.47 \\
0551$-$226 N  &  0.800  &  0.709   &  45.0 & 0.19  &   14.3  &  0.59 \\
0551$-$226 S  &  0.800  &  0.730   &  20.0 & 0.29  &    6.6  &  0.89 \\
       &         &          &       &        &         &         \\
0937$-$250 N  &  0.900  &  0.812   &  39.5 & 0.36  &   13.1  &  1.09 \\
0937$-$250 S  &  0.900  &  0.932   &  22.0 & 0.40  &    8.7  &  1.02 \\
0938$-$205 N  &  0.371  &  0.441   &  66.0 & 0.17  &   23.1  &  0.49 \\
0938$-$205 S  &  0.371  &  0.462   &  75.5 & 0.15  &   28.0  &  0.39 \\
0947$-$249 N  &  0.854  &  1.147   &  40.0 & 0.30  &   20.8  &  0.58 \\
0947$-$249 S  &  0.854  &  0.841   &  39.0 & 0.15  &   14.4  &  0.41 \\
0955$-$283 E  &  0.800  &  0.696   &  47.5 & 0.23  &   14.7  &  0.75 \\
0955$-$283 W  &  0.800  &  0.856   &  39.5 & 0.29  &   16.0  &  0.72 \\
1022$-$250 E  &  0.340  &  0.605   &  58.5 & 0.15  &   30.8  &  0.29 \\
1022$-$250 W  &  0.340  &  0.661   &  47.0 & 0.13  &   26.8  &  0.22 \\
       &         &          &       &        &         &         \\
1023$-$226 N  &  0.586  &  0.744   &  52.5 & 0.17  &   24.2  &  0.37 \\
1023$-$226 S  &  0.586  &  0.552   &  61.5 & 0.12  &   19.7  &  0.38 \\
1025$-$229 N  &  0.309  &  0.313   &  80.5 & 0.09  &   19.8  &  0.38 \\
1025$-$229 S  &  0.309  &  0.296   & 134.0 & 0.05  &   30.2  &  0.20 \\
1026$-$202 N  &  0.566  &  0.632   &  54.5 & 0.16  &   21.4  &  0.41 \\
1026$-$202 S  &  0.566  &  0.735   &  35.0 & 0.21  &   16.3  &  0.45 \\
1029$-$233 E  &  0.611  &  0.866   &  49.0 & 0.19  &   25.5  &  0.36 \\
1029$-$233 W  &  0.611  &  0.693   &  33.5 & 0.16  &   13.7  &  0.38 \\
1052$-$272 N  &  1.103  &  1.173   &  19.5 & 0.30  &    7.9  &  0.75 \\
1052$-$272 S  &  1.103  &  0.996   &  20.5 & 0.39  &    6.8  &  1.16 \\
\end{tabular}
\end{table}

\begin{table}
\begin{tabular}{l l l r l r l }
       &         &          &       &        &         &         \\
       &         &          &       &        &         &         \\
Source &  z      & B$_{eq}$& Age   & Vel    & Age$^*$ & Vel$^*$ \\
Name   &         &  nT      & Myr   & 0.1c   &  Myr    &  0.1c   \\
       &         &          &       &        &         &         \\
1107$-$218 E  &  1.500  &  1.278   &  14.5 & 0.49  &    4.2  &  1.67 \\
1107$-$218 W  &  1.500  &  1.542   &  22.5 & 0.42  &    8.4  &  1.12 \\
1107$-$227 N  &  2.000  &  2.371   &   8.5 & 1.04  &    3.4  &  2.58 \\
1107$-$227 S  &  2.000  &  2.139   &   8.5 & 0.96  &    3.0  &  2.71 \\
1126$-$290 N  &  0.410  &  0.562   &  59.0 & 0.13  &   25.8  &  0.30 \\
1126$-$290 S  &  0.410  &  0.514   &  94.0 & 0.12  &   37.2  &  0.30 \\
1224$-$208 N  &  1.500  &  1.188   &  18.5 & 0.52  &    4.8  &  2.01 \\
1224$-$208 S  &  1.500  &  0.965   &  16.5 & 0.49  &    3.1  &  2.59 \\
1226$-$297 N  &  0.749  &  0.615   &  36.5 & 0.05  &   10.3  &  0.17 \\
1226$-$297 S  &  0.749  &  1.221   &  15.5 & 0.41  &    9.4  &  0.67 \\
       &         &          &       &        &         &         \\
1232$-$249 N  &  0.352  &  1.024   &  34.0 & 0.20  &   25.6  &  0.26 \\
1232$-$249 S  &  0.352  &  0.820   &  35.0 & 0.18  &   23.2  &  0.27 \\
1247$-$290 N  &  0.770  &  0.991   &  26.0 & 0.25  &   12.8  &  0.50 \\
1247$-$290 S  &  0.770  &  0.841   &  36.5 & 0.34  &   15.1  &  0.82 \\
1257$-$230 N  &  1.109  &  1.756   &   2.5 & 2.93  &    1.5  &  4.85 \\
1257$-$230 S  &  1.109  &  1.842   &   8.0 & 1.27  &    5.0  &  2.03 \\
2035$-$203 E  &  0.516  &  0.832   &  22.0 & 0.22  &   12.3  &  0.39 \\
2035$-$203 W  &  0.516  &  0.701   &  25.5 & 0.15  &   12.1  &  0.31 \\
2040$-$236 E  &  0.704  &  1.270   &  15.0 & 0.31  &    9.8  &  0.48 \\
2040$-$236 W  &  0.704  &  0.642   &  31.0 & 0.19  &   10.0  &  0.58 \\
       &         &          &       &        &         &         \\
2042$-$293 N  &  1.900  &  1.237   &  16.5 & 0.68  &    2.9  &  3.91 \\
2042$-$293 S  &  1.900  &  1.055   &  16.0 & 0.41  &    2.1  &  3.06 \\
2045$-$245 N  &  0.730  &  0.857   &  44.5 & 0.14  &   19.8  &  0.33 \\
2045$-$245 S  &  0.730  &  0.759   &  30.0 & 0.20  &   11.6  &  0.51 \\
2118$-$266 E  &  0.343  &  0.483   &  60.0 & 0.10  &   24.7  &  0.25 \\
2118$-$266 W  &  0.343  &  0.409   &  66.5 & 0.06  &   22.2  &  0.17 \\
2132$-$235 N  &  0.810  &  0.727   &  37.0 & 0.28  &   12.0  &  0.85 \\
2132$-$235 S  &  0.810  &  0.864   &  25.0 & 0.20  &   10.1  &  0.50 \\
2137$-$279 N  &  0.640  &  0.875   &  27.0 & 0.44  &   13.7  &  0.87 \\
2137$-$279 S  &  0.640  &  0.814   &  19.5 & 0.17  &    9.2  &  0.36 \\
       &         &          &       &        &         &         \\
2213$-$283 E  &  0.946  &  1.040   &  14.0 & 0.39  &    5.9  &  0.93 \\
2213$-$283 W  &  0.946  &  1.252   &  16.5 & 0.29  &    8.5  &  0.56 \\
2311$-$211 E  &  0.434  &  1.159   &  14.5 & 0.45  &   11.0  &  0.60 \\
2311$-$211 W  &  0.434  &  0.751   &  32.0 & 0.15  &   18.1  &  0.26 \\
2325$-$213 N  &  0.580  &  1.024   &  12.5 & 0.69  &    7.8  &  1.11 \\
2325$-$213 S  &  0.580  &  1.086   &  13.0 & 0.45  &    8.4  &  0.69 \\
2338$-$290 N  &  0.446  &  0.529   &  35.5 & 0.08  &   13.6  &  0.20 \\
2338$-$290 S  &  0.446  &  0.479   &  41.0 & 0.23  &   13.9  &  0.68 \\
2348$-$235 N  &  0.952  &  1.042   &  13.0 & 0.31  &    5.5  &  0.72 \\
2348$-$235 S  &  0.952  &  1.116   &  14.0 & 0.59  &    6.4  &  1.29 \\
\end{tabular}
\end{table}

Following Myers \& Spangler (1985), a quantity X$_\circ$ is defined as $$
X_\circ = {C_2^2 \over C_1} \nu B^3t^2, $$ which is an indicator of the
synchrotron age of the source. The constants $C_2 \ \ {\rm and}\ \ C_1$
are defined in Pacholczyk (1970), $B$ is the magnetic field strength,
$\nu$ is the radio frequency and $t$ is the time since the electrons were
accelerated.  From the above expression, the age $ t_{_{Myr}}$ is given by
$$ t_{_{Myr}} = {\sqrt{X_\circ} \over 2.98\times 10^{-2}[(1 +
z)\nu_{_{GHz}}]^{1/2} B^{3/2}_{-5}}. $$ \noindent Here $ t_{_{Myr}}$ is
the time in millions of years, $B_{-5}$ is the magnetic field strength in
units of 10$^{-5}$ G, $\nu_{_{GHz}}$ is the observing frequency in GHz and
$z$, the redshift of the source.  The quantity X$_\circ$ can be obtained
by assuming that the electron spectrum evolves due to synchrotron
radiation losses over a time $t$.

\begin{figure}
\vbox{
\vspace{-1.0 in}
\hspace{0.05 in}
\psfig{figure=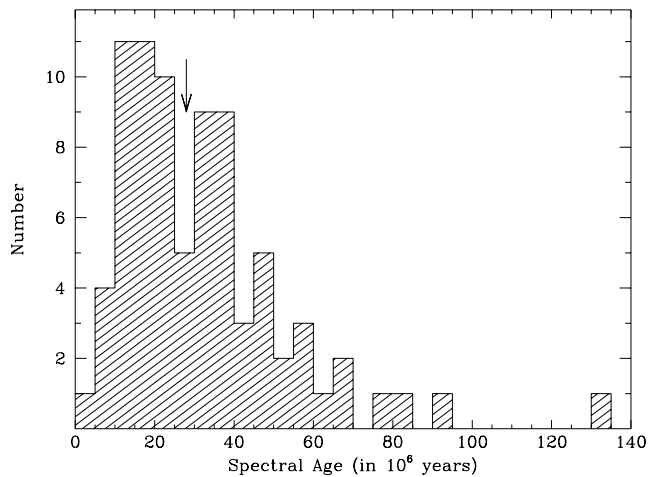,height=3.5in}
\vspace{0.15 in}
}
\caption{Distribution of the spectral ages of Molonglo radio sources due
to synchrotron radiation losses. The median value is shown by an arrow. }
\end{figure}

The spectral ages and the speeds of separations were calculated using the
magnetic field B$_{eq}$ estimated from the minimum energy arguments from the
L- and C-band observations. The 15 GHz observations were not used in the 
estimates of spectral ages because of the lack of detection of the lobe
emission at this frequency. The
age estimates were not done for northern lobe of 0346$-$297 because of poor 
signal to noise ratio, for northern lobe of 0454$-$220 because the spectral 
index variations were within the measurement errors and for the radio galaxy 
1358$-$214 which does not have the C-band data.
The distribution of the spectral ages is presented in Figure 6. The spectral
ages, equipartition magnetic field strengths, and the velocity estimates
are presented in Table 6, which is arranged as follows.  Columns 1 \& 2:
Source name and its redshift; column 3: the equipartition magnetic field
in nT, columns 4 \& 5: age in 10$^6$ yr and speed of separation in units
of 0.1c, Column 6 \& 7: same as Columns 4 \& 5, but after taking into
account the radiation losses due to inverse-Compton scattering with the
microwave background radiation. The synchrotron ages range from 2.5
$\times 10^6$ years to 1.34 $\times 10^8$ years with a median value of
about 2.8 $\times 10^7$ years, and the speed of separation is in the range
0.0046c to 0.29c with a median value of 0.025c. If we take account of the
energy losses due to inverse-Compton scattering with the microwave
background radiation, the ages are in the range of 1.5 $\times 10^6$ years
to 3.7 $\times 10^7$ years with a median value of about 1.2 $\times 10^7$
years while, the speed of separation is in the range 0.015c to 0.48c with
a median value of 0.058c.

\begin{figure*}
\hbox{
\psfig{figure=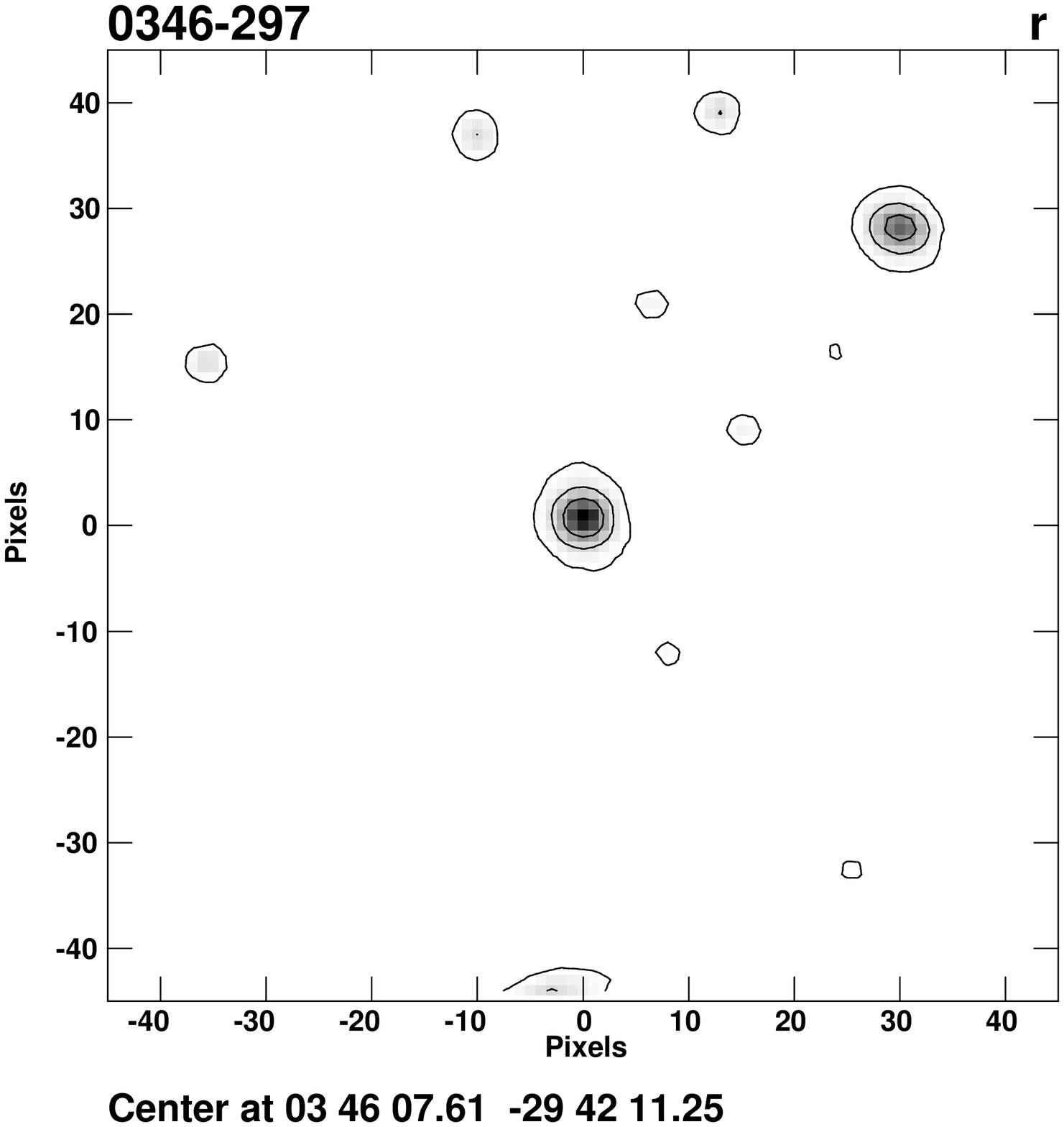,height=2.25in}
\hspace{0.25in}
\psfig{figure=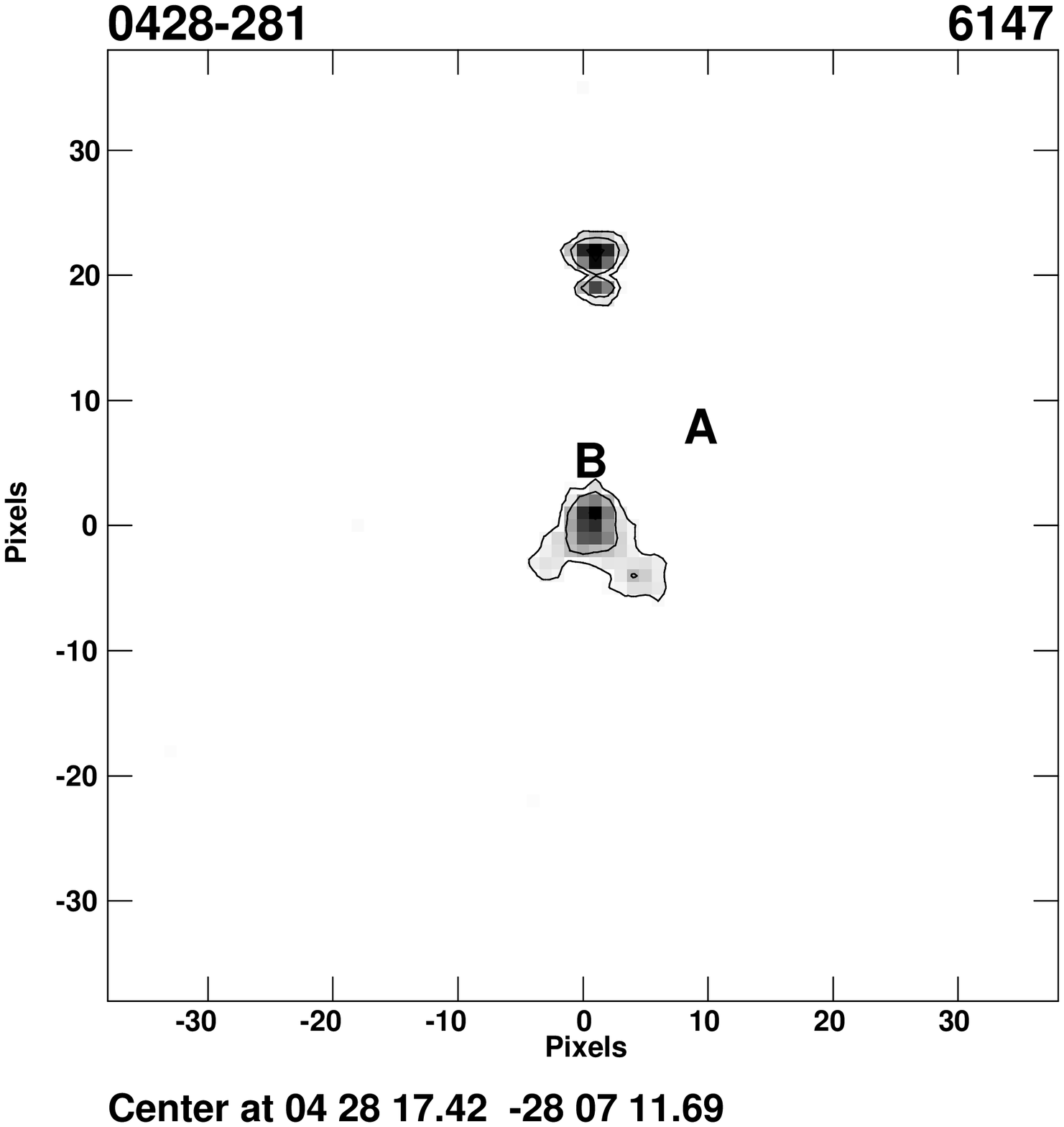,height=2.25in}
\hspace{0.25in}
\psfig{figure=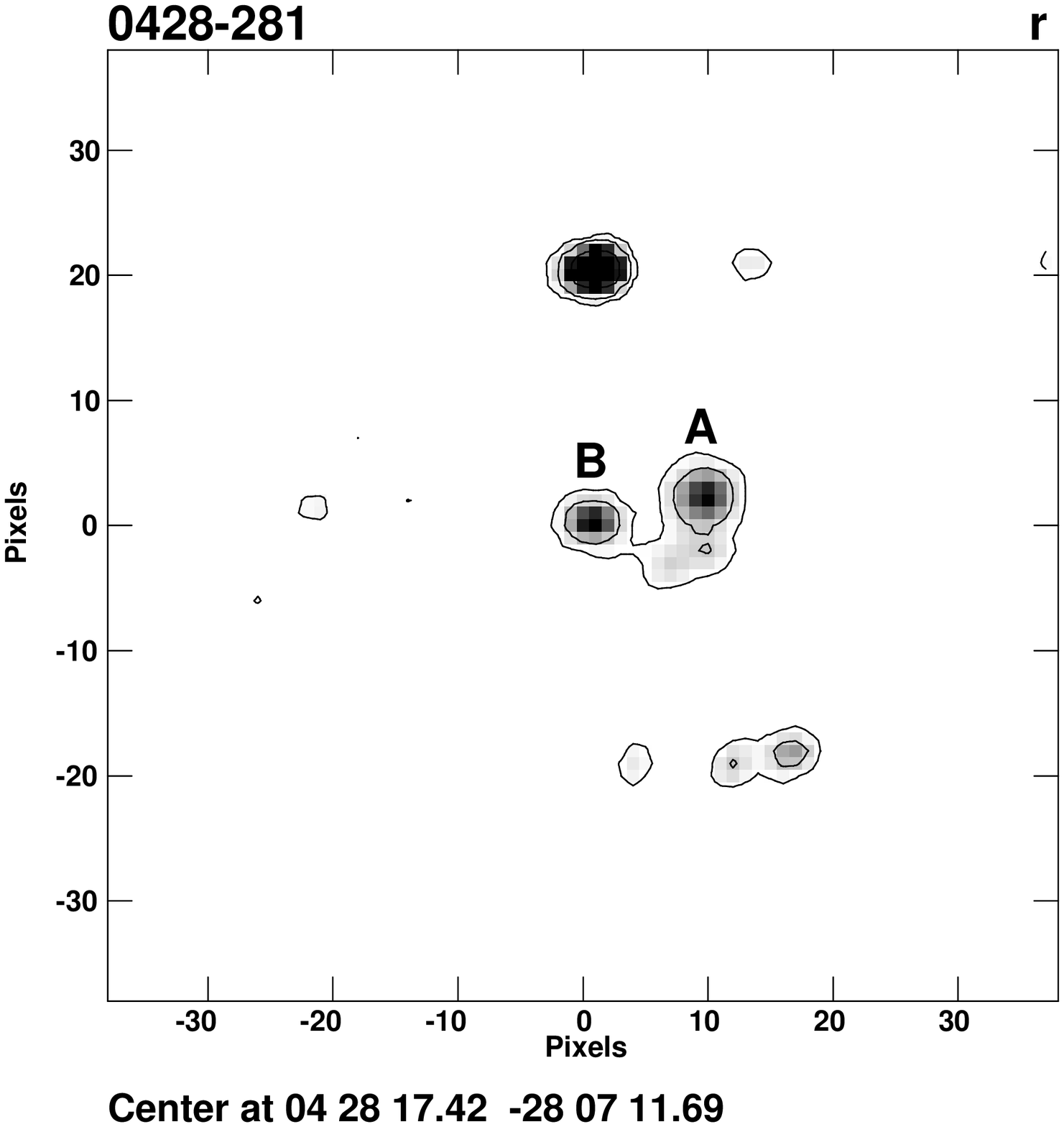,height=2.25in}
}
\vspace{-0.05in}
\hbox{
\psfig{figure=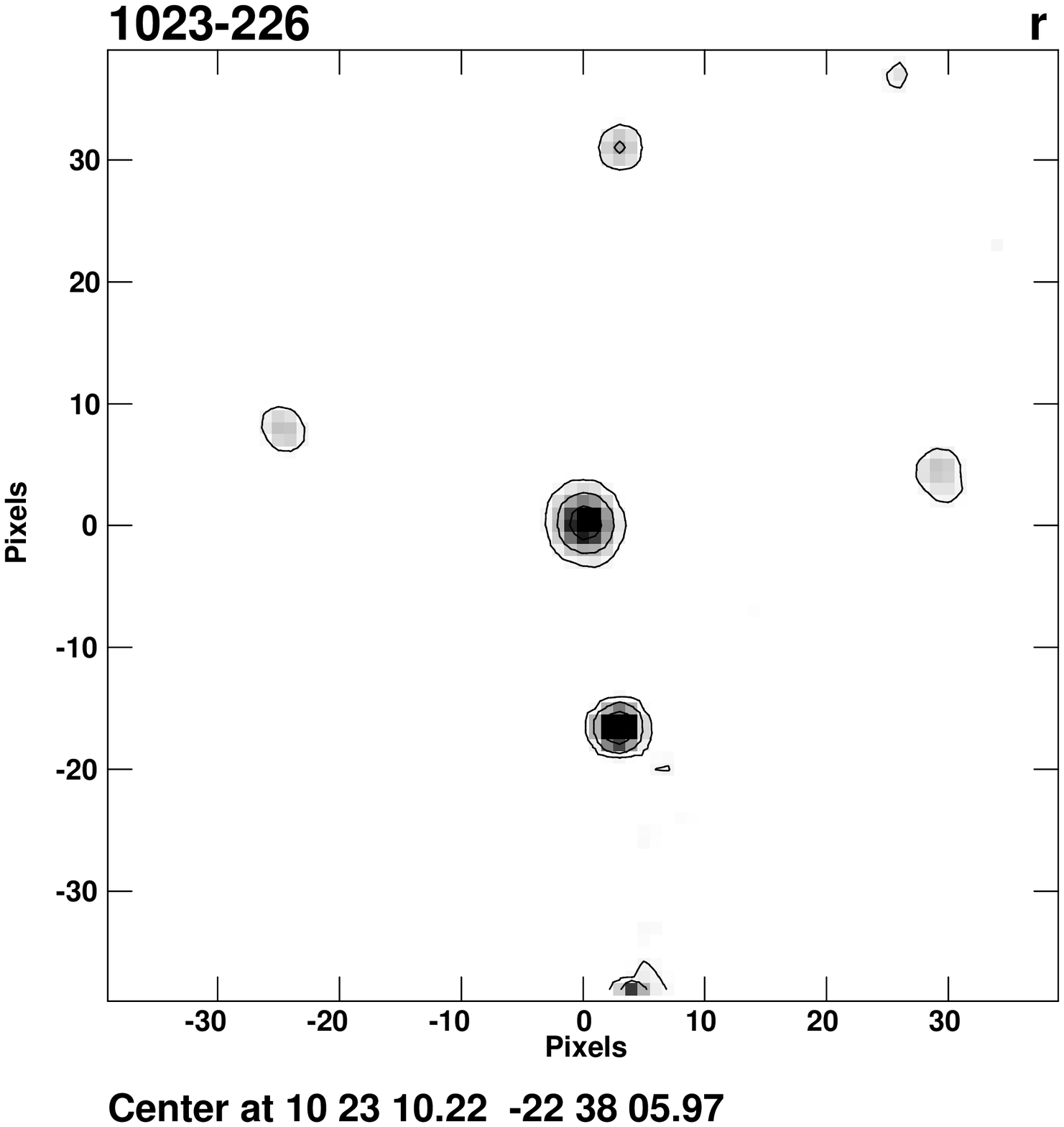,height=2.25in}
\hspace{0.25in}
\psfig{figure=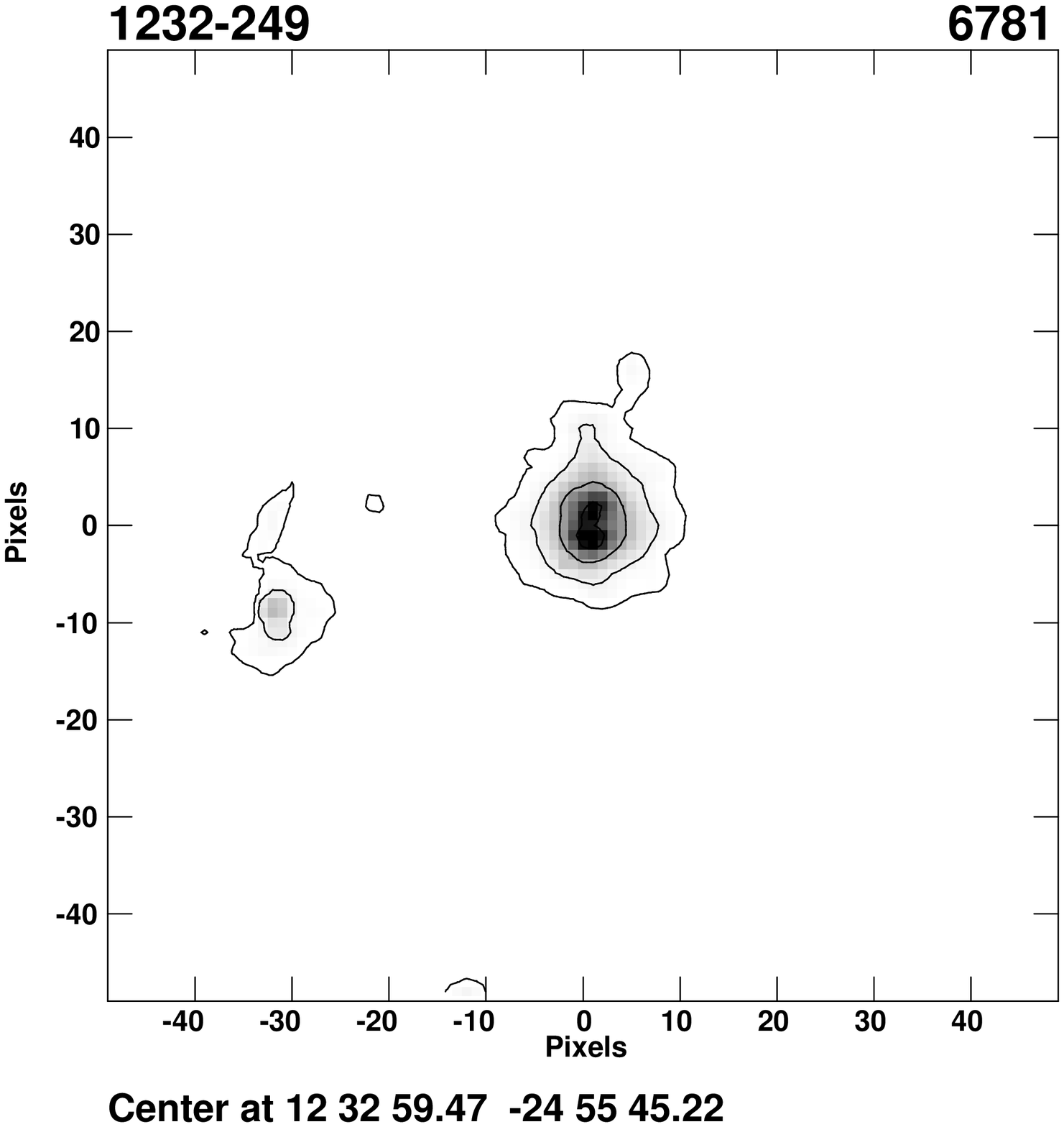,height=2.25in}
\hspace{0.25in}
\psfig{figure=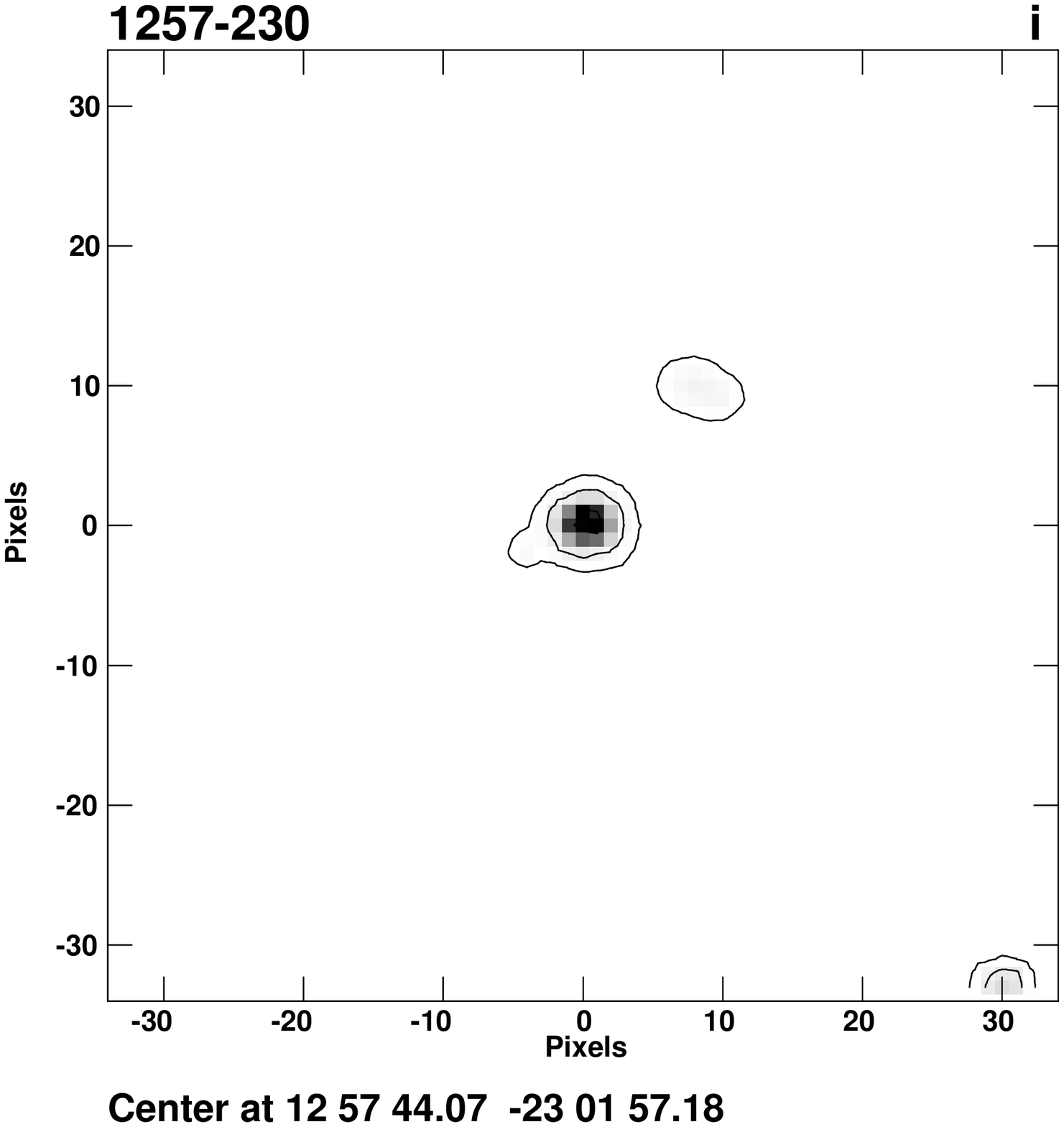,height=2.25in}
}
\vspace{-0.05in}
\caption[Narrow-and broad-band images of radio galaxies and quasars.]
{Narrow- and broad-band images of radio galaxies and quasars. The x and 
y axes are in pixels, each pixel corresponding to 0.52 arcsec. The source name
is in the top left corner of the image and the observing filter in the 
top right corner. Each window has a width of 300 kpc. The centres in
B1950 co-ordinates are given at the bottom of every image. North is to the
top and east is to the left.}
\end{figure*}

\section{Results from the optical observations}

The narrow- and broad-band images of the radio galaxies and quasars are
presented in Figure 7 for the objects with significant extension of
emission-line gas and evidence of galaxies in the broad-band images for
signs of possible interaction with the radio source. The optical images 
for all the sources are presented in Ishwara-Chandra (1999.) We now compare 
the distribution of emission-line
gas and galaxies along the radio source axis and correlate these with the
symmetry parameters of the radio source. Four galaxies and two quasars
have narrow-band observation, and significant extension of emission-line
gas is seen in one quasar (1232$-$249) and possibly in one galaxy
(0428$-$281). In the case of this weak-cored quasar, the emission-line gas
is extended within about 20$^\circ$ of the radio source axis, consistent
with the alignment effect noted in high-redshift radio galaxies (Chambers
et al. 1987; McCarthy et al. 1987). In the case of the radio galaxy
0428$-$281, the
extension is possibly due to the companion object. The
radio lobe of 1232$-$249 lie well outside the emission-line gas extension, 
and do not show any significant depolarization between
1.4 and 5 GHz (IC98).

We have also looked for evidence of galaxies in the broad-band images for
signs of possible interaction with the radio source. In the broad-band
images, three galaxies (0346$-$297, 0428$-$281, 1023$-$226) and possibly
in one quasar (1257$-$230), a galaxy is seen close to the radio axis. The
arm-length ratios for 0428$-$281 and 1023$-$226 are consistent with
slowing down of the jet by possible interaction with the galaxy. In the
radio galaxy 3C34 where Best, Longair \& R\"{o}ttgering (1997) have
presented evidence of jet-cloud interaction at a distance of 120 kpc from
the nucleus of the parent optical galaxy, the primary hotspot on the side
of the interaction is closer to the nucleus than the one on the opposite
side. This is consistent with the these two sources. However, in the case
of 0346$-$297, where an optical galaxy is seen towards the north-west lobe
of the radio source, the radio lobe on this side is weaker and longer
compared to the lobe on the opposite side. For the quasar 1257$-$230, a
faint galaxy-like object is seen towards the northern lobe, within
10$^\circ$ of the radio source axis. This radio lobe is shorter, but
weaker compared to the lobe on the opposite side. We comment below briefly
on the above objects.

\subsection*{Notes on individual sources}

\noindent {\bf 0346$-$297:} This galaxy has a very asymmetric radio
morphology. Almost the entire radio flux density of the source is due to
south-eastern lobe. There is no evidence of extended emission-line gas. In
the broad-band $r$ image a galaxy is seen about 135 kpc from the radio
galaxy towards the north-west within about 10$^\circ$ of the radio source
axis. The radio lobe on this side is weaker and farther away ($r_\theta
\sim$ 1.5)  from the host galaxy compared to the lobe on the opposite
side.

\noindent {\bf 0428$-$281:} There appears to be weak extension of the
emission-line gas distribution. An optical galaxy (B) is seen physically
connected to the galaxy (A) identified by McCarthy et al. (1996) .
However, in the continuum subtracted emission-line image, the galaxy A
disappears completely and the galaxy B appears bright. The astrometric
position of galaxy B is RA 04h 28m 17.44s and DEC $-28^\circ$ 07$^\prime$
11.$^{\prime\prime}$9 (B1950), while that of galaxy A is RA 04h 28m 17.09s
and DEC $-28^\circ$ 07$^\prime$ 10.$^{\prime\prime}$9 (B1950). The
position of the possible radio core seen in the 8 GHz image is RA = 04h
28m 17.33s and DEC $-28^\circ$ 07$^\prime$ 11.$^{\prime\prime}$20 (B1950),
which is within about 1.$^{\prime\prime}$5 of the position of B. If the
radio core is confirmed, this is likely to be the correct identification.
The galaxy A lies about 40 kpc west of B within 10$^\circ$ of the radio
source axis. Extended emission is seen in the narrow band image after 
subtracting the continuum emission of the galaxy A.

\noindent {\bf 1023$-$226:} In the  
broad-band image a faint optical galaxy is seen within about 20$^\circ$ of
the radio axis, at a distance of about 110 kpc towards north-west. The
arm-length ratio of the lobe on this direction is 0.75 and the brightness
ratio is 1.8.

\noindent {\bf 1232$-$249:} The emission-line gas in this quasar is
elongated towards the north, within about 20$^\circ$ of the radio source
axis. This lobe is farther from the nucleus than the opposite lobe. Both
the lobes do not show depolarization between 1.4 and 5GHz. Lower radio
frequency polarization observations along with deeper emission-line
imaging will be useful in understanding the relationship between the
emission-line gas distribution and possible depolarization. 

\noindent {\bf 1257$-$230:} This quasar also has only broad-band
observations which shows a faint galaxy within about 10$^\circ$ of the
northern radio lobe. The radio lobe on this side is shorter and fainter.

\section{Concluding remarks}

We have presented the results of our radio observations at the X and U
bands, and the optical narrow- and broad-band images. We summarise the
main conclusions in this section.

\begin{enumerate}

\item Considering the objects where the difference between the spectral
indices and depolarization values of the lobes on opposite sides is larger
than the errors, the Liu-Pooley relation appears stronger for sources of
smaller linear size, with no signficant dependence on the redshift of the
objects. This relationship which shows that the lobe with the flatter
spectral index is less depolarized is similar for both radio galaxies and
quasars, suggesting that in addition to orientation effects intrinsic
differences also play a role.

\item In the brightess ratio - armlength ratio diagram derived from our
observations with an angular resolution of about 5 arcsec, there is a
significant trend for the nearer component to be brighter, especially for
galaxies. This component also tends to be more depolarized (cf. IC98),
suggesting that the nearer component is moving through a denser medium
leading to larger dissipation of energy and more depolarization. The
tendency for the nearer component to be brighter is weaker when considers
the flux density of the hotspots from our higher resolution observations
with an angular resolution of about 1 arcsec. The effects of mild
relativisitic motion of the hotspots appear to be more noticeable in the
high-resolution images.

\item The ages of the relativistic electrons in the lobes due to radiative
losses range from 1.5 $\times 10^6$ to 3.7 $\times 10^7$ years with a
median value of about 1.2 $\times 10^7$ years while, the speed of
separation is in the range 0.015c to 0.48c with a median value of 0.058c.

\item A significant extension of emission-line gas was seen in one quasar
1232$-$249 and possibly in one galaxy 0428$-$281.  In the case of the
quasar, the emission line gas is extended within about 20$^\circ$ of the
radio source axis, while in the radio galaxy 0428$-$281, it may be related
to an apparent companion galaxy.

\item We have identified four cases where there appears to be a galaxy
projected within about 20$^\circ$ of the position angle of the source. In
two of these (0428$-$281 and 1023$-$226) their arm-length and brightness
ratios are consistent with slowing down of the jet by possible interaction
with the galaxy. 

\end{enumerate} 

\section*{Acknowledgments} We thank Yogesh Wadadekar for
introducing us to IRAF. We are indebted to an anonymous 
referee for very helpful and critical comments on the paper.
The work by WvB was performed under the auspices
of the U.S. Department of Energy by University of California Lawrence
Livermore National Laboratory under contract No.  W-7405-Eng-48. The
National Radio Astronomy Observatory is a facility of the National Science
Foundation operated under co-operative agreement by Associated
Universities Inc. We thank the staff of the Very Large Array for the
observations.  This research has made use of the NASA/IPAC extragalactic
database (NED) which is operated by the Jet Propulsion Laboratory,
Caltech, under contract with the National Aeronautics and Space
Administration.

{}

\begin{table}
\subsection*{Appendix: The sample of sources }
\begin{tabular}{cclrcrr }
Source & Id   & z & S$_{1365}$ & P$_{1365}$ & LAS & LLS \\
 Name  &      &      & mJy  & W/Hz/sr   & $''$& kpc \\
           &   &        &        &         &         &     \\
0017-207 & Q & 0.545 &  467 &  25.76 &  96   &  705 \\
0058-229 & Q & 0.706 &  396 &  25.95 &  63   &  505 \\
0133-266 & Q & 1.53  &  348 &  26.61 &  53   &  454 \\
0137-263 & G & 1.1   &  509 &  26.51 &  76   &  655 \\
0148-297 & G & 0.41  & 2778 &  26.28 & 136   &  883 \\
0325-260 & G & 0.638 &  286 &  25.70 &  57   &  445 \\
0346-297 & G & 0.413 &  620 &  25.65 & 127   &  831 \\
0428-281 & G & 0.65  &  956 &  26.25 &  62   &  482 \\
0437-244 & Q & 0.84  &  459 &  26.19 & 125   & 1040 \\
0454-220 & Q & 0.533 & 1993 &  26.35 &  83   &  608 \\
           &   &        &        &         &         &     \\
0551-226 & G & 0.8   &  330 &  26.00 &  53   &  438 \\
0937-250 & G & 0.9   &  445 &  26.28 &  63   &  533 \\
0938-205 & G & 0.371 &  444 &  25.39 &  88   &  543 \\
0947-249 & G & 0.854 & 1487 &  26.75 &  69   &  575 \\
0955-283 & G & 0.8   &  470 &  26.14 &  75   &  620 \\
1022-250 & G & 0.34  &  338 &  25.20 &  52   &  305 \\
1023-226 & G & 0.586 &  298 &  25.64 &  58   &  436 \\
1025-229 & Q & 0.309 &  489 &  25.28 & 186   & 1041 \\
1026-202 & G & 0.566 &  664 &  25.95 &  61   &  456 \\
1029-233 & G & 0.611 &  403 &  25.80 &  68   &  521 \\
           &   &        &        &         &         &     \\
1052-272 & Q & 1.103 &  553 &  26.56 &  73   &  630 \\
1107-218 & G & 1.5   &  302 &  26.53 &  54   &  460 \\
1107-227 & G & 2.0   &  781 &  27.40 &  64   &  527 \\
1126-290 & G & 0.41  & 1086 &  25.89 & 104   &  676 \\
1224-208 & G & 1.5   &  268 &  26.57 &  61   &  517 \\
1226-297 & Q & 0.749 &  435 &  26.00 &  65   &  525 \\
1232-249 & Q & 0.352 & 1988 &  25.97 & 108   &  648 \\
1247-290 & Q & 0.77  &  696 &  26.27 &  58   &  470 \\
1257-230 & Q & 1.109 &  788 &  26.65 &  51   &  441 \\
1358-214 & G & 0.5   &  317 &  25.34 &  88   &  629 \\
           &   &        &        &         &         &     \\
2035-203 & Q & 0.516 &  752 &  25.87 &  63   &  457 \\
2040-236 & Q & 0.704 &  428 &  25.89 &  54   &  431 \\
2042-293 & G & 1.9   &  414 &  26.99 &  66   &  547 \\
2045-245 & G & 0.73  &  639 &  26.21 &  76   &  615 \\
2118-266 & G & 0.343 &  403 &  25.24 &  86   &  510 \\
2132-235 & G & 0.81  &  333 &  26.02 &  54   &  444 \\
2137-279 & G & 0.64  &  455 &  25.93 &  58   &  447 \\
2213-283 & Q & 0.946 &  820 &  26.53 &  58   &  491 \\
2311-222 & G & 0.434 &  965 &  25.85 &  79   &  528 \\
2325-213 & G & 0.58  &  986 &  26.17 &  75   &  566 \\
           &   &        &        &         &         &     \\
2338-290 & Q & 0.446 &  427 &  25.51 &  72   &  489 \\
2348-235 & G & 0.952 &  503 &  26.34 &  67   &  570 \\
\end{tabular}
\end{table}
\end{document}